\documentclass[aps,prb,preprint,superscriptaddress]{revtex4-1}
\usepackage{amsmath,amsfonts,amssymb}

\begin{document}


\title{On the effective potential of Duru-Kleinert path integrals}


\author{Seiji Sakoda}
\email[]{sakoda@nda.ac.jp}
\affiliation{Department of Applied Physics, National Defense Academy,
Hashirimizu\\
Yokosuka city, Kanagawa 239-8686, Japan}


\date{\today}

\begin{abstract}
We propose a new method to evaluate the effective potential in the path integral for the fixed-energy amplitude as well as for the pseudotime evolution kernel in the formalism by Duru and Kleinert. Restriction to the postpoint or the prepoint prescriptions in formulating time sliced path integrals is avoided by leaving off the use of expectation values for correction terms. This enables us to consider an arbitrary ordering prescription and to examine the ordering dependence of the effective potential.
To investigate parameter dependences, we introduce the ordering parameter $\alpha$ in addition to the splitting parameter $\lambda$ in the formulation of the time sliced path integral. The resulting path integrals are found to be independent of the ordering parameter although the explicit dependence, given by a contribution proportional to $(1-2\lambda)^{2}$, on the splitting parameter remains. As an application, we check the relationship between path integrals for the radial oscillator and the radial Coulomb system in arbitrary dimensions.
\end{abstract}


\maketitle

\section{Introduction}
The exact solution of the path integral of the hydrogen atom by Duru and Kleinert\cite{DK} opened a new area in the path integral formalism. The essence of their method is the combination of a path-dependent time reparametrization and a compensating coordinate transformation. After their elegant work, many attempts\cite{RingwoodDevreese,BlanchardSirugue,Inomata81,HoInomata,Inomata,Steiner,PakSokmen,ChetouaniChetouani,YoungDeWitt-Morette,Kleinert87,CastrigianoStaerk} have been made to clarify their formalism. 
Details of the Coulomb path integral and applications of the Duru-Kleinert(DK) method can be found in the comprehensive textbook\cite{KleinertBook} by Kleinert. Consideration from gauge invariance viewpoint on the DK method was first given by Fujikawa to make it clear that the essence of the DK transformation can be viewed as the special choice of the gauge fixing condition for a system with invariance under reparametrization of time\cite{Fujikawa,Kleinert98}.
An application of Fujikawa's approach, which is based on the view point of the Jacobi's principle of least action, was found by the present author in formulating an exactly solvable path integral of one-dimensional Coulomb system\cite{Sakoda} which does not posses any oscillator coordinates to be transformed to.

As the original paper by Duru and Kleinert related the hydrogen atom to the isotropic oscillator in four dimensions, their method converts a path integral of a dynamical system into another one. In the course of this transmutation of dynamical systems, we need to evaluate the complicated kinetic term to obtain additional potential terms in addition to the usual quadratic kinetic term for the transformed system. The evaluation of this effective potential seems to be done in the literature\cite{KleinertBook} by replacing higher order terms of the difference $\Delta q$, where $q$ is related to the original coordinate $x$ by $x=h(q)$, with their expectation values against the Gaussian weight for the measure $d\Delta q$ within the time sliced path integral. Of course, as was shown by Feynman for path integrals over Cartesian coordinates\cite{Feynman48} also by McLaughlin and Schulman even for those in curved spaces\cite{McLaughlinSchulman}, it is useful to evaluate the higher order terms of $\Delta q$ by their expectation values against the Gaussian weight $\exp\{-(\Delta q)^{2}/(2\epsilon)\}$ with the measure $d\Delta q$ for infinitesimally small pseudotime interval $\epsilon$. 
The use of this technique in the DK transformed path integral will also produce a correct result for a time sliced path integral. However, if the order of performing integrations is restricted in a specific manner as is seen at the bottom of page 533 of Ref.\onlinecite{KleinertBook}, the measure of such path integrals defined by a product of $d\Delta q$'s may be unpleasant. For instance, if we try to performe the WKB approximation or other similar methods to evaluate such path integrals, the restriction in the order of integrations will immediately give rise to obstacles. Furthermore, as we will see in the next section, the derivation of the effective potential and the proof of its independence of the so called splitting parameter $\lambda$, introduced soon below, is possible only for path integrals formulated by postpoint or prepoint prescriptions. Therefore,  for path integrals in the DK method, the invention of a novel technique which does not rely on such an inconvenient measure will be awaiting a detailed study. 
Our strategy in this paper to achieve the aim explained above is to cease making use of the measure defined by a product of $d\Delta q$'s and to formulate path integrals with the measure given by a  product of $dq$'s instead.
In doing so, the procedure of making a change of variables, the evaluation of the Jacobian in particular, in the DK transformation may naturally become different from the one formulated as multiple integrals against $d\Delta q$'s. 
It is therefore important to study the detail of the procedure of DK transformation from our own viewpoint since this possible difference may cause a change in the effective potential of the DK transformed path integral.
To this aim, the splitting parameter $\lambda$ for the regulating functions $f_{l}(x)=\{f(x)\}^{1-\lambda}$ and $f_{r}(x)=\{f(x)\}^{\lambda}$ which yield time reparametrization $dt=f(x)ds$ in the DK formalism will be a good test. We shall try to formulate the time sliced path integral whose Hamiltonian to be parametrized by the ordering parameter $\alpha$ although the prescription given in Chapter 14 of Ref.\onlinecite{KleinertBook} is restricted to the postpoint form according to the measure $d\Delta x$. In addition to the splitting parameter $\lambda$, the ordering parameter $\alpha$ will also be a touchstone for our method to be developed in this paper by checking the dependence in resulting path integrals on these parameters.

For our purpose it is meaningless to formulate path integrals without time slicing since we deal with quite formal and delicate issues. To formulate a time sliced path integral by keeping good connections with the operator formalism, we need a set of completeness or the resolution of unity for eigenvectors of some nice operators in addition to that of the position operator. Completeness of eigenvectors of the momentum operator will be assumed in this paper excepting in the section we consider the application of our method to the radial path integral.

The paper is organized as follows: in section 2 we first make a brief review of the derivation of the effective potential by Kleinert. The limitation of the method shown there will be made clear to motivate the invention of a novel formulation. We will then consider the conversion of the complicated kinetic term into the effective potential both for time sliced path integrals of the Feynman kernel and the fixed-energy amplitude in section 3. The application of our method developed in section 2 to radial path integrals will be shown in section 4. The final section will be devoted to the conclusion. Details of the procedure to convert the complicated kinetic term into the effective potential will be given in the appendix.

\section{Derivation of the effective potential by Kleinert}

For a Hamiltonian
\begin{equation}
\label{eq:kleinert01}
H=\frac{\ 1\ }{\ 2m\ }p^{2}+V(x),
\end{equation}
Duru-Kleinert method defines a new one given by
\begin{equation}
\label{eq:kelinert02}
H_{E}=f_{l}(x)(H-E)f_{r}(x)
\end{equation}
and formulates a path integral for the fixed-energy amplitude
\begin{equation}
\label{eq:kelinert03}
G(x,x{'};E)\equiv
\langle x\vert\frac{\ 1\ }{\ H-E\ }\vert x{'}\rangle=
\frac{\ 1\ }{\ \hbar\ }\int_{0}^{\infty}\!d\sigma\,
f_{r}(x)f_{l}(x{'})K_{E}(x,x{'};\sigma),
\end{equation}
where the Feynman kernel $K_{E}(x,x{'};\sigma)$ is defined by
\begin{equation}
\label{eq:kleinert04}
K_{E}(x,x{'};\sigma)=
\langle x\vert e^{-H_{E}\sigma/\hbar}\vert x{'}\rangle
\end{equation}
with imaginary pseudotime $\sigma$. Here $f_{l}(x)f_{r}(x)=f(x)$($f(x)>0$) and $f(x)$ is related to $h(q)$ in $x=h(q)$, that describes the change of variables from $x$ to $q$ below, by $\{h{'}(q)\}^{2}=f(x)$.

Introducing $\epsilon_{\sigma}=\sigma/N$ and keeping always the limit $N\to\infty$ in mind,
we can express the Feynman kernel above as a time sliced path integral:
\begin{equation}
\label{eq:kleinert05}
\begin{aligned}
K_{E}(x,x{'};\sigma)=&\int\!\prod_{i=1}^{N-1}dx_{i}\,
\int\!\prod_{j=1}^{N}\frac{\ dp_{j}\ }{\ 2\pi\hbar\ }\,
\exp\left[\frac{\ i\ }{\ \hbar\ }\sum_{k=1}^{N}p_{k}\Delta x_{k}-\right.\\
&\left.
\frac{\ \epsilon_{\sigma}\ }{\ \hbar\ }\sum_{k=1}^{N}
\left\{\frac{\ 1\ }{\ 2m\ }f_{l}(x_{k})f_{r}(x_{k-1})p_{k}^{2}+
f_{l}(x_{k})f_{r}(x_{k-1})\{V(x^{(\alpha)}_{k})-E\}\right\}\right],
\end{aligned}
\end{equation}
where for the argument of the potential $V(x)$ we adopted $\alpha$-ordering
$x^{(\alpha)}_{j}=(1-\alpha)x_{j}+\alpha x_{j-1}=x_{j}-\alpha\Delta x_{j}$.
By carrying out integrations with respect to $p_{j}$'s, we obtain
\begin{equation}
\label{eq:kleinert06}
\begin{aligned}
&K_{E}(x,x{'};\sigma)=
\frac{\ 1\ }{\ \sqrt{f_{l}(x_{N})f_{r}(x_{0})\,}\ }
\left(\frac{\ m\ }{\ 2\pi\hbar\epsilon_{\sigma}\ }\right)^{N/2}
\int\!\prod_{i=1}^{N-1}\frac{\ dx_{i}\ }{\ \sqrt{f(x_{i})\,}\ }\\
&\times\exp\left[-\sum_{j=1}^{N}\left\{
\frac{\ m\ }{\ 2\hbar\epsilon_{\sigma}\ }
\left(
\frac{\ \Delta x_{j}\ }{\ \sqrt{f_{l}(x_{j})f_{r}(x_{j-1})\,}\ }
\right)^{2}
-\frac{\ \epsilon_{\sigma}\ }{\ \hbar\ }
f_{l}(x_{j})f_{r}(x_{j-1})\{V(x^{(\alpha)}_{j})-E\}\right\}\right].
\end{aligned}
\end{equation}

Substitution of \eqref{eq:kleinert06} into \eqref{eq:kelinert03} leads us to a time sliced path integral:
\begin{equation}
\label{eq:kleinert07}
\begin{aligned}
&G(x,x{'};E)=\frac{\ 1\ }{\ \hbar\ }\int_{0}^{\infty}\!d\sigma\,
\frac{\ f_{r}(x_{N})f_{l}(x_{0})\ }{\ \sqrt{f_{l}(x_{N})f_{r}(x_{0})\,}\ }
\left(\frac{\ m\ }{\ 2\pi\hbar\epsilon_{\sigma}\ }\right)^{N/2}
\int\!\prod_{i=1}^{N-1}\frac{\ dx_{i}\ }{\ \sqrt{f(x_{i})\,}\ }\\
&\times\exp\left[-\sum_{j=1}^{N}\left\{
\frac{\ m\ }{\ 2\hbar\epsilon_{\sigma}\ }
\left(
\frac{\ \Delta x_{j}\ }{\ \sqrt{f_{l}(x_{j})f_{r}(x_{j-1})\,}\ }
\right)^{2}
-\frac{\ \epsilon_{\sigma}\ }{\ \hbar\ }
f_{l}(x_{j})f_{r}(x_{j-1})\{V(x^{(\alpha)}_{j})-E\}\right\}\right]
\end{aligned}
\end{equation}
for the fixed-energy amplitude.

As far as the requirement $f_{l}(x)f_{r}(x)=f(x)$ is fulfilled,
$f_{l}(x)$ and $f_{r}(x)$ are arbitrary, we may, however, here follow Kleinert\cite{KleinertBook} to assume that they are parameterized by the splitting parameter $\lambda$:
\begin{equation}
\label{eq:kleinert08}
f_{l}(x)=\{f(x)\}^{1-\lambda},\
f_{r}(x)=\{f(x)\}^{\lambda}.
\end{equation}
By rewriting
\begin{equation}
\label{eq:kleinert09}
\frac{\ f_{r}(x_{N})f_{l}(x_{0})\ }{\ \sqrt{f_{l}(x_{N})f_{r}(x_{0})\,}\ }=
\{f(x_{N})\}^{(3\lambda-1)/2}\{f(x_{0})\}^{1-3\lambda/2}
\end{equation}
as
\begin{equation}
\label{eq:kleinert10}
\frac{\ f_{r}(x_{N})f_{l}(x_{0})\ }{\ \sqrt{f_{l}(x_{N})f_{r}(x_{0})\,}\ }=
\{f(x_{N})f(x_{0})\}^{1/4}
\left\{\frac{\ f(x_{0})\ }{\ f(x_{N})\ }\right\}^{\gamma/2},\quad
\gamma=\frac{\ 3\ }{\ 2\ }(1-2\lambda),
\end{equation}
and making use of an identity
\begin{equation}
\label{eq:kleinert11}
\left\{\frac{\ f(x_{0})\ }{\ f(x_{N})\ }\right\}^{\gamma/2}=
\prod_{j=1}^{N}\left\{\frac{\ f(x_{j-1})\ }{\ f(x_{j})\ }\right\}^{\gamma/2},
\end{equation}
we arrange the form of the prefactor to be
\begin{equation}
\label{eq:kleinert12}
\frac{\ f_{r}(x_{N})f_{l}(x_{0})\ }{\ \sqrt{f_{l}(x_{N})f_{r}(x_{0})\,}\ }=
\{f(x_{N})f(x_{0})\}^{1/4}
\prod_{j=1}^{N}
\left\{\frac{\ h{'}(q_{j-1})\ }{\ h{'}(q_{j})\ }\right\}^{\gamma},
\end{equation}
where use has been made of the relation $\{h{'}(q)\}^{2}=f(x)$.
If we treat the path integral in the postpoint prescription, it will be natural here to consider the Taylor expansion of the ratio $h{'}(q_{j-1})/h{'}(q_{j})$ around $q_{j}$ in terms $\Delta q_{j}\equiv q_{j}-q_{j-1}$ as
\begin{equation}
\label{eq:referee01}
	\frac{\ h{'}(q_{j-1})\ }{\ h{'}(q_{j})\ }=1-
	b_{2}(q_{j})\Delta q_{j}+
	\frac{1}{\ 2\ }b_{3}(q_{j})(\Delta q_{j})^{2}+\cdots\ ,
\end{equation}
where we have defined
\begin{equation}
\label{eq:kleinert14}
b_{2}(q)\equiv\frac{\ h{''}(q)\ }{\ h{'}(q)\ },\
b_{3}(q)\equiv\frac{\ h{'''}(q)\ }{\ h{'}(q)\ },\
\dots\ .
\end{equation}Then the exponentiation of the factor $\{h{'}(q_{j-1})/h{'}(q_{j})\}^{\gamma}$ combined together with another factor from the measure will make contribution to the effective potential of the path integral.
When we formulate, however, the path integral in the $\alpha$-ordering prescription, defined by $q^{(\alpha)}_{j}=(1-\alpha)q_{j}+\alpha q_{j-1}$, we need to rewrite the effective potential obtained in the postpoint prescription to be expressed in terms of $q^{(\alpha)}_{j}$. 
Therefore it is convenient to consider here the Taylor series of the ratio above around $q^{(\alpha)}_{j}$ as
\begin{equation}
\label{eq:kleinert13}
\frac{\ h{'}(q_{j-1})\ }{\ h{'}(q_{j})\ }=
1-b_{2}(q^{(\alpha)}_{j})\Delta q_{j}+
\left[\alpha\{b_{2}(q^{(\alpha)}_{j})\}^{2}-
\left(\alpha-\frac{\ 1\ }{\ 2\ }\right)b_{3}(q^{(\alpha)}_{j})\right]
(\Delta q_{j})^{2}+\cdots.
\end{equation}
We will make use of this expression in the next section in formulating the path integral within a generalized ordering prescription.

In Chapter 14 of Ref.\onlinecite{KleinertBook}, $\alpha=0$ is chosen so that we can rewrite the measure in \eqref{eq:kleinert07} as
\begin{equation}
\label{eq:kleinert15}
\prod_{i=1}^{N-1}\frac{\ dx_{i}\ }{\ \sqrt{f(x_{i})\,}\ }=
\sqrt{\frac{\ f(x_{N})\ }{\ f(x_{1})\ }\,}
\prod_{i=1}^{N-1}\frac{\ d\Delta x_{i+1}\ }{\ \sqrt{f(x_{i+1})\,}\ }.
\end{equation}
Then, discarding irrelevant contributions, the measure can be further rewritten as
\begin{equation}
\label{eq:kelinert16}
\prod_{i=1}^{N-1}\frac{\ dx_{i}\ }{\ \sqrt{f(x_{i})\,}\ }=
\prod_{i=1}^{N-1}\frac{\ d\Delta x_{i+1}\ }{\ \sqrt{f(x_{i+1})\,}\ }
\prod_{j=1}^{N}\left\{\frac{\ f(x_{j})\ }{\ f(x_{j-1})\ }\right\}^{1/2}.
\end{equation}
By combining this result together with the prefactor, we obtain
\begin{equation}
\label{eq:kleinert17}
\frac{\ f_{r}(x_{N})f_{l}(x_{0})\ }{\ \sqrt{f_{l}(x_{N})f_{r}(x_{0})\,}\ }
\prod_{i=1}^{N-1}\frac{\ dx_{i}\ }{\ \sqrt{f(x_{i})\,}\ }=
\{f(x_{N})f(x_{0})\}^{1/4}
\prod_{i=1}^{N-1}\frac{\ d\Delta x_{i+1}\ }{\ \sqrt{f(x_{i+1})\,}\ }
\prod_{j=1}^{N}
\left\{\frac{\ h{'}(q_{j-1})\ }{\ h{'}(q_{j})\ }\right\}^{\gamma_{0}},
\end{equation}
where
\begin{equation}
\label{eq:kleinert18}
\gamma_{0}\equiv\frac{\ 1\ }{\ 2\ }(1-6\lambda)=\gamma-1
\end{equation}
has been introduced. 
Making a change of variables from $\Delta x_{j}$'s to $\Delta q_{j}$'s produces a Jacobian
\begin{equation}
\label{eq:kleinert19}
J_{0}=\prod_{j=2}^{N}\sqrt{f(x_{j})\,}\left\{
1-b_{2}(q_{j})\Delta q_{j}+
\frac{\ 1\ }{\ 2\ }b_{3}(q_{j})(\Delta q_{j})^{2}-+\cdots\right\}
\end{equation}
while the kinetic term in the exponent of \eqref{eq:kleinert07} can be expanded as 
\begin{equation}
\label{eq:kleinert20}
\begin{aligned}
&\frac{\ m\ }{\ 2\hbar\epsilon_{\sigma}\ }
\frac{\ (\Delta x_{j})^{2}\ }{\ f_{l}(x_{j})f_{r}(x_{j-1})\ }=
\frac{\ m\ }{\ 2\hbar\epsilon_{\sigma}\ }
\left[\vphantom{\frac{\ 1\ }{\ 3\ }}
(\Delta q_{j})^{2}+(2\lambda-1)b_{2}(q_{j})(\Delta q_{j})^{3}+\right.\\
&\left.
\hphantom{\frac{\ m\ }{\ 2\hbar\epsilon_{\sigma}\ }}
\left\{\left(\frac{\ 1\ }{\ 3\ }-\lambda\right)b_{3}(q_{j})+
\left(\frac{\ 1\ }{\ 4\ }+\lambda(2\lambda-1)\right)\{b_{2}(q_{j})\}^{2}
\right\}(\Delta q_{j})^{4}+\cdots\right]
\end{aligned}
\end{equation}
to be expanded further into the perturbative series against the Gaussian weight
\begin{equation}
\label{eq:kleinert21}
\exp\left\{-\frac{\ m\ }{\ 2\hbar\epsilon_{\sigma}\ }
\sum_{j=1}^{N}(\Delta q_{j})^{2}\right\}
\end{equation}
with the product measure $\prod_{i=1}^{N-1}d\Delta q_{i+1}$.
Upon these preparations, it is shown that the effective potential is independent of $\lambda$ and given by
\begin{equation}
\label{eq:kleinert22}
V_{\mathrm{eff}}(q)=
\frac{\ \hbar^{2}\ }{\ 2m\ }\left[
\frac{\ 3\ }{\ 4\ }
\left\{\frac{\ h{''}(q)\ }{\ h{'}(q)\ }\right\}^{2}-
\frac{\ 1\ }{\ 2\ }\frac{\ h{'''}(q)\ }{\ h{'}(q)\ }
\right]
\end{equation}
within the framework of the postpoint prescription.
It must be emphasized here that the order of integrals with respect to $\Delta q_{j}$'s is specified to be done from $j=N$ to $j=2$ in the successive manner while keeping $q_{j+1}$ fixed at each step. We have already take it into account in the evaluation of the Jacobian above. If we wish to be free from this restriction, we must express $q_{j}$ as
\begin{equation}
\label{eq:kleinert23}
q_{j}=q_{N}-\sum_{k=j+1}^{N}\Delta q_{k}
=q_{0}+\sum_{k=1}^{j}\Delta q_{k}
\end{equation}
by taking a constraint
\begin{equation}
\sum_{j=1}^{N}\Delta q_{j}=q_{N}-q_{0}
\end{equation}
into account. This will cause, however, a change in the Jacobian, hence in the effective potential, through derivatives of coefficients in the power series of $\Delta q_{j}$.

For understanding the limitation of the derivation above, it will be helpfull to consider the prepoint prescription instead.
To fit this prescription, we first rewrite the measure as
\begin{equation}
\label{eq:kleinert24}
\prod_{i=1}^{N-1}\frac{\ dx_{i}\ }{\ \sqrt{f(x_{i})\,}\ }=
\sqrt{\frac{\ f(x_{0})\ }{\ f(x_{N-1})\ }\,}
\prod_{i=1}^{N-1}\frac{\ d\Delta x_{i}\ }{\ \sqrt{f(x_{i-1})\,}\ }=
\prod_{i=1}^{N-1}\frac{\ d\Delta x_{i}\ }{\ \sqrt{f(x_{i-1})\,}\ }
\prod_{j=1}^{N}\left\{\frac{\ f(x_{j-1})\ }{\ f(x_{j})\ }\right\}^{1/2}
\end{equation}
by discarding irrelevant terms. The product of the prefactor with the measure is then expressed as
\begin{equation}
\label{eq:kleinert25}
\frac{\ f_{r}(x_{N})f_{l}(x_{0})\ }{\ \sqrt{f_{l}(x_{N})f_{r}(x_{0})\,}\ }
\prod_{i=1}^{N-1}\frac{\ dx_{i}\ }{\ \sqrt{f(x_{i})\,}\ }=
\{f(x_{N})f(x_{0})\}^{1/4}
\prod_{i=1}^{N-1}\frac{\ d\Delta x_{i}\ }{\ \sqrt{f(x_{i-1})\,}\ }
\prod_{j=1}^{N}
\left\{\frac{\ h{'}(q_{j-1})\ }{\ h{'}(q_{j})\ }\right\}^{\gamma_{1}},
\end{equation}
where
\begin{equation}
\label{eq:kleinert26}
\gamma_{1}\equiv\frac{\ 3\ }{\ 2\ }(1-\lambda)=\gamma+1.
\end{equation}
The Jacobian for switching from $\Delta x_{j}$'s to $\Delta q_{j}$'s is given by
\begin{equation}
\label{eq:kleinert27}
J_{1}=\prod_{j=1}^{N-1}\sqrt{f(x_{j-1})\,}\left\{
1+b_{2}(q_{j-1})\Delta q_{j}+
\frac{\ 1\ }{\ 2\ }b_{3}(q_{j-1})(\Delta q_{j})^{2}+\cdots\right\}
\end{equation}
and the suitable expansion of the kinetic term will be
\begin{equation}
\label{eq:kleinert28}
\begin{aligned}
&\frac{\ m\ }{\ 2\hbar\epsilon_{\sigma}\ }
\frac{\ (\Delta x_{j})^{2}\ }{\ f_{l}(x_{j})f_{r}(x_{j-1})\ }=
\frac{\ m\ }{\ 2\hbar\epsilon_{\sigma}\ }
\left[\vphantom{\frac{\ 1\ }{\ 3\ }}
(\Delta q_{j})^{2}+(2\lambda-1)b_{2}(q_{j-1})(\Delta q_{j})^{3}+\right.\\
&\left.
\hphantom{\frac{\ m\ }{\ 2\hbar\epsilon_{\sigma}\ }}
\left\{\left(\lambda-\frac{\ 2\ }{\ 3\ }\right)b_{3}(q_{j-1})+
\left(\frac{\ 1\ }{\ 4\ }+(\lambda-1)(2\lambda-1)\right)\{b_{2}(q_{j-1})\}^{2}
\right\}(\Delta q_{j})^{4}+\cdots\right]
\end{aligned}
\end{equation}
for this prescription. 
Again, it should be noted that integrals with respect to $\Delta q_{j}$ for this case must be carried out from $j=1$ to $j=N-1$ in this order by keeping $q_{j-1}$ fixed at each step. 

The perturbative expansion
\begin{equation}
\label{eq:kleinert29}
\begin{aligned}
&\exp\left\{-\frac{\ 1\ }{\ 2\epsilon\ }
\frac{\ (\Delta x_{j})^{2}\ }{\ f_{l}(x_{j})f_{r}(x_{j-1})\ }\right\}=
\exp\left\{-\frac{\ 1\ }{\ 2\epsilon\ }(\Delta q_{j})^{2}\right\}
\left[1-\frac{\ 1\ }{\ 2\epsilon\ }\left\{
\vphantom{\frac{1}{4}}
(2\lambda-1)b_{2}(q_{j-1})(\Delta q_{j})^{3}+\right.\right.\\
&\hphantom{\frac{\ (\Delta x_{j})^{2}\ }{\ f_{l}(x_{j})f_{r}(x_{j-1})\ }}
\left.\left.
\left\{\left(\frac{\ 1\ }{\ 4\ }+(\lambda-1)(2\lambda-1)\right)
\{b_{2}(q_{j-1})\}^{2}+
\left(\lambda-\frac{\ 2\ }{\ 3\ }\right)b_{3}(q_{j-1})\right\}
(\Delta q_{j})^{4}\right\}+\right.\\
&\hphantom{\frac{\ (\Delta x_{j})^{2}\ }{\ f_{l}(x_{j})f_{r}(x_{j-1})\ }}
\left.
\frac{\ 1\ }{\ 8\epsilon^{2}\ }(2\lambda-1)^{2}\{b_{2}(q_{j-1})\}^{2}
(\Delta q_{j})^{6}+\cdots\right],\quad
\epsilon\equiv\frac{\ \hbar\epsilon_{\sigma}\ }{\ m\ },
\end{aligned}
\end{equation}
and the Jacobian as well as the prefactor will be, by dropping odd powers of $\Delta q_{j}$, then brought together into
\begin{equation}
\label{eq:kleinert30}
\begin{aligned}
&1+\frac{\ \gamma_{1}\ }{\ 2\ }\left[
(\gamma_{1}+1)\{b_{2}(q_{j-1})\}^{2}-b_{3}(q_{j-1})\right](\Delta q_{j})^{2}-\\
&\frac{\ 1\ }{\ 2\epsilon\ }\left[
\left(\frac{\ 1\ }{\ 4\ }+(\lambda-\gamma_{1}-1)(2\lambda-1)\right)\{b_{2}(q_{j-1})\}^{2}+
\left(\lambda-\frac{\ 2\ }{\ 3\ }\right)b_{3}(q_{j-1})\right\}
(\Delta q_{j})^{4}+\\
&\frac{\ 1\ }{\ 8\epsilon^{2}\ }(2\lambda-1)^{2}\{b_{2}(q_{j-1})\}^{2}
(\Delta q_{j})^{6}+\cdots\,,
\end{aligned}
\end{equation} 
to be evaluated as an expectation value against the Gaussian weight \eqref{eq:kleinert21} by utilizing the formula
\begin{equation}
\label{eq:kleinert31}
\frac{\ 1\ }{\ \sqrt{2\pi\epsilon\,}\ }\int_{-\infty}^{\infty}\!
e^{-x^{2}/(2\epsilon)}x^{2n}\,dx=(2n-1)!!\epsilon^{n}.
\end{equation}
We thus obtain the same effective potential as given by \eqref{eq:kleinert22} and prove that the effective potential is independent of $\lambda$ even in the framework of the prepoint prescription.
Changes in the Jacobian, the prefactor and in the perturbative expansion of the kinetic term upon the transition from postpoint to prepoint prescription totally cancel to result in the same effective potential.

We are thus convinced that, not only in the framework of the postpoint but also in that of prepoint, we can prove the $\lambda$-independence of the effective potential or the uniqueness of \eqref{eq:kleinert22} for the fixed-energy amplitude in the Duru-Kleinert method. Since $x_{0}(=h(q_{0})=x{'})$ and $x_{N}(=h(q_{N})=x)$ are fixed, however, there is no room for considering the ordering prescription other than $\alpha=0$ or $\alpha=1$; we cannot fix $x^{(\alpha)}_{N}=x_{N}-\alpha\Delta x_{N}$ when we carry out the integration with respect to $\Delta x_{N}$ except the case $\alpha=0$, for example. Therefore the derivation of the effective potential and the proof of its $\lambda$-independence is possible only for these two special prescriptions. 
In order to discuss the effective potential for an arbitrary choice of the ordering parameter, we should not stick to the measure $d\Delta x_{j}$ or $d\Delta q_{j}$ which plays the role in the perturbative evaluation of the correction terms. We have to, therefore, develop a new method for this purpose and leave away from the use of expectation values for correction terms against the Gaussian weight.

\section{A new derivation of the effective potential}

In this section we discuss a new formulation of the time sliced path integral for the Duru-Kleinert method to find the effective potential for an arbitrary ordering prescription parametrized by $\alpha$.
To begin with, let us consider the Feynman kernel \eqref{eq:kleinert06}:
\begin{equation}
\label{eq:kernel-a2}
\begin{aligned}
&K_{E}(x,x{'};\sigma)=
\frac{\ 1\ }{\ \sqrt{f_{l}(x_{N})f_{r}(x_{0})\,}\ }
\left(\frac{\ m\ }{\ 2\pi\hbar\epsilon_{\sigma}\ }\right)^{N/2}
\int\!\prod_{i=1}^{N-1}dq_{i}\\
&\times\exp\left[-\sum_{j=1}^{N}\left\{
\frac{\ m\ }{\ 2\hbar\epsilon_{\sigma}\ }
\left\{
\Delta q_{j}R(q^{(\alpha)}_{j},\Delta q_{j})
\right\}^{2}
-\frac{\ \epsilon_{\sigma}\ }{\ \hbar\ }
f_{l}(x_{j})f_{r}(x_{j-1})\{V(x^{(\alpha)}_{j})-E\}\right\}\right]
\end{aligned}
\end{equation}
in which $R(q^{(\alpha)}_{j},\Delta q_{j})$ is defined by
\begin{equation}
\frac{\ \Delta x_{j}\ }{\ \sqrt{f_{l}(x_{j})f_{r}(x_{j-1})\,}\ }=
\Delta q_{j}R(q^{(\alpha)}_{j},\Delta q_{j}),\
q^{(\alpha)}_{j}=(1-\alpha)q_{j}+\alpha q_{j-1},\
\Delta q_{j}=q_{j}-q_{j-1}
\end{equation}
and use has been made of the relation $x=h(q)$ as well as $\sqrt{f(x)\,}=h{'}(q)$.
Since the potential term is not essential for the argument below, we leave it as it is for a while and focus attention on conversion of higher order terms in $\Delta q_{j}R(q^{(\alpha)}_{j},\Delta q_{j})$, which will be given by a power series of $\Delta q_{j}$'s, into an effective potential. 

To this end, we first need to find the explicit form of $R(q^{(\alpha)}_{j},\Delta q_{j})$ above.
Recalling the relations $f_{l}(x)=\{f(x)\}^{1-\lambda}$, $f_{r}(x)=\{f(x)\}^{\lambda}$ and $x=h(q)$, we can rewrite $\Delta x_{j}/\sqrt{f_{l}(x_{j})f_{r}(x_{j-1})\,}$ as
\begin{equation}
\frac{\ \Delta x_{j}\ }{\ \sqrt{f_{l}(x_{j})f_{r}(x_{j-1})\,}\ }=
\frac{\ h(q_{j})-h(q_{j-1})\ }
{\ \{h{'}(q_{j})\}^{1-\lambda}\{h{'}(q_{j-1})\}^{\lambda}\ }.
\end{equation}
By expressing $x_{j}$ and $x_{j-1}$ by $q^{(\alpha)}_{j}$ and $\Delta q_{j}$ as
\begin{equation}
x_{j}=h(q_{j})=h(q^{(\alpha)}_{j}+\alpha\Delta q_{j}),\
x_{j-1}=h(q_{j-1})=h(q^{(\alpha)}_{j}-\beta\Delta q_{j})\
(\beta=1-\alpha),
\end{equation}
we expand them into the power series of $\Delta q_{j}$ to find
\begin{equation}
\Delta x_{j}=h{'}(q^{(\alpha)}_{j})\Delta q_{j}\left\{1+
\frac{\ 1\ }{\ 2\ }(\alpha^{2}-\beta^{2})b_{2}(q^{(\alpha)}_{j})\Delta q_{j}+
\frac{\ 1\ }{\ 6\ }(\alpha^{3}+\beta^{3})b_{3}(q^{(\alpha)}_{j})(\Delta q_{j})^{2}+
\cdots\right\}.
\end{equation}
In the same way, we find
\begin{equation}
\begin{aligned}
&\{h{'}(q_{j})\}^{-1+\lambda}\{h{'}(q_{j-1})\}^{-\lambda}=
\{h{'}(q^{(\alpha)}_{j})\}^{-1}
\left[\vphantom{\frac{1}{2}}
1+(\lambda-\alpha)b_{2}(q^{(\alpha)}_{j})\Delta q_{j}+\right.\\
&\left.
\frac{\ 1\ }{\ 2\ }\left[\left\{(\alpha-\lambda)^{2}+
\lambda(1-2\alpha)+\alpha^{2}\right\}\{b_{2}(q^{(\alpha)}_{j})\}^{2}-
\left\{\alpha^{2}+\lambda(1-2\alpha)\right\}b_{3}(q^{(\alpha)}_{j})
\right](\Delta q_{j})^{2}+\cdots\right].
\end{aligned}
\end{equation}
These are combined into the series expansion
\begin{equation}
\begin{aligned}
&\frac{\ \Delta x_{j}\ }{\ \sqrt{f_{l}(x_{j})f_{r}(x_{j-1})\,}\ }=
\Delta q_{j}\left[
1-\frac{\ 1\ }{\ 2\ }(1-2\lambda)b_{2}(q^{(\alpha)}_{j})\Delta q_{j}+\right.\\
&\left.
\frac{\ 1\ }{\ 2\ }\left[\vphantom{\frac{1}{2}}
\left\{(1-2\lambda)\alpha+\lambda^{2}\right\}
\{b_{2}(q^{(\alpha)}_{j})\}^{2}-
\left\{(1-2\lambda)\alpha+\lambda-\frac{\ 1\ }{\ 3\ }
\right\}b_{3}(q^{(\alpha)}_{j})\right](\Delta q_{j})^{2}+\cdots\right]
\end{aligned}
\end{equation}
which yields the explicit form of $R(q^{(\alpha)}_{j},\Delta q_{j})$:
\begin{equation}
\label{eq:coeffs01}
\begin{aligned}
R(q^{(\alpha)}_{j},\Delta q_{j})=&
1+a_{2}(q^{(\alpha)}_{j})\Delta q_{j}+a_{3}(q^{(\alpha)}_{j})(\Delta q_{j})^{2}+\cdots,\\
a_{2}(q)=&
-\frac{\ 1\ }{\ 2\ }(1-2\lambda)b_{2}(q),\\
a_{3}(q)=&
\frac{\ 1\ }{\ 2\ }\left[\vphantom{\frac{1}{2}}
\left\{(1-2\lambda)\alpha+\lambda^{2}\right\}
\{b_{2}(q)\}^{2}-
\left\{(1-2\lambda)\alpha+\lambda-\frac{\ 1\ }{\ 3\ }
\right\}b_{3}(q)\right],\ \cdots.
\end{aligned}
\end{equation}

If we define
\begin{equation}
\label{eq:change01}
q{'}_{j}\equiv q_{j}+\sum_{k=1}^{j}
\{R(q^{(\alpha)}_{k},\Delta q_{k})-1\}\Delta q_{k}\ 
(j=1,\,2,\,\dots,\,N),\
q{'}_{0}\equiv q_{0},
\end{equation}
to find
\begin{equation}
\Delta q{'}_{j}\equiv q{'}_{j}-q{'}_{j-1}=
\Delta q_{j}R(q^{(\alpha)}_{j},\Delta q_{j}),
\end{equation}
we can rewrite \eqref{eq:kernel-a2} as
\begin{equation}
\label{eq:kernel-a1}
\begin{aligned}
K_{E}(x,x{'};\sigma)=&
\frac{\ 1\ }{\ \sqrt{f_{l}(x_{N})f_{r}(x_{0})\,}\ }
\left(\frac{\ m\ }{\ 2\pi\hbar\epsilon_{\sigma}\ }\right)^{N/2}
\int\!\prod_{i=1}^{N-1}dq_{i}\\
&\times
\exp\left[-\sum_{j=1}^{N}\left\{\frac{\ m\ }{\ 2\hbar\epsilon_{\sigma}\ }
(\Delta q{'}_{j})^{2}+
\frac{\ \epsilon_{\sigma}\ }{\ \hbar\ }
f_{l}(x_{j})f_{r}(x_{j-1})\{V(x^{(\alpha)}_{j})-E\}\right\}\right].
\end{aligned}
\end{equation}
As is shown in the appendix the matrix defined by $\partial q_{i}/\partial q{'}_{j}$($i,\,j=1,\,2,\,\dots,\,N-1$) is triangular, its determinant is given by the product of its diagonal elements.
We thus find, through the calculation given in the appendix, that the Jacobian of the change of variables $q_{j}\mapsto q{'}_{j}$($j=1,\,2,\,\dots,\,N-1$) above
can be exponentiated as
\begin{equation}
\label{eq:jacobian02}
\exp\left[
-\sum_{j=1}^{N}\left\{C_{1}(q{'}^{(\alpha)}_{j})\Delta q{'}_{j}+
\frac{\ 1\ }{\ 2\ }C_{2}(q{'}^{(\alpha)}_{j})(\Delta q{'}_{j})^{2}\right\}
\right],
\end{equation}
where
\begin{equation}
\label{eq:coeffs02}
C_{1}(q)\equiv
2a_{2}(q),\quad
C_{2}(q)\equiv
2\left\{3a_{3}(q)-
4\{a_{2}(q)\}^{2}+
(1-\alpha)a_{2}{'}(q)\right\},
\end{equation}
by keeping the exponent correct up to $O(\epsilon_{\sigma})$. As a result the quadratic form of $\Delta q{'}_{j}$ in the exponent of \eqref{eq:kernel-a1} is modified by additional terms from the Jacobian. We can, however, rearrange the quadratic form by introducing another set of new variables $\xi_{j}$'s which differ from $q{'}_{j}$'s by higher order terms in $\epsilon_{\sigma}$.
By this change of variables, the quadratic form of $\Delta q{'}_{j}$ in the exponent of \eqref{eq:kernel-a1} is converted into the effective potential in addition to the standard kinetic term. We shall postpone the detail of this conversion in the appendix and here we just make use of the result obtained there to rewrite the time sliced path integral \eqref{eq:kernel-a1} as
\begin{equation}
\label{eq:kernel-a3}
\begin{aligned}
&K_{E}(x,x{'};\sigma)=\frac{\ 1\ }{\ \sqrt{f_{l}(x_{N})f_{r}(x_{0})\,}\ }
\left(\frac{\ m\ }{\ 2\pi\hbar\epsilon_{\sigma}\ }\right)^{N/2}
\int_{-\infty}^{\infty}\!\prod_{i=1}^{N-1}d\xi_{i}\,
\exp\left[-\sum_{j=1}^{N}\frac{\ m\ }{\ 2\hbar\epsilon_{\sigma}\ }
(\Delta \xi_{j})^{2}\right]\\
&\times
\exp\left[
-\frac{\ \epsilon_{\sigma}\ }{\ \hbar\ }\sum_{j=1}^{N}\left[
f_{l}(x_{j})f_{r}(x_{j-1})\{V(x^{(\alpha)}_{j})-E\}+\right.\right.\\
&\left.\left.\hphantom{\times
\exp\left[-\frac{\ \epsilon_{\sigma}\ }{\ \hbar\ }\right]}
\frac{\ \hbar^{2}\ }{\ 2m\ }\left\{
\{C_{1}(\xi^{(\alpha)}_{j})\}^{2}+
C_{2}(\xi^{(\alpha)}_{j})
+2(1-\alpha)C_{1}{'}(\xi^{(\alpha)}_{j})
\right\}\right]\vphantom{\sum_{j=1}^{N}}\right],
\end{aligned}
\end{equation}
where $\xi_{j}$ has been introduced by
\begin{equation}
\label{eq:q'2xi}
\xi_{j}=\sqrt{1+\epsilon G(\eta_{j})\,}\eta_{j},\quad
\eta_{j}=
q{'}_{j}+\sum_{k=1}^{j}
\frac{\ \epsilon C_{1}(q{'}^{(\alpha)}_{k})\ }
{\ 1+\epsilon C_{2}(q{'}^{(\alpha)}_{k})\ },\quad
\epsilon\equiv\frac{\ \hbar\epsilon_{\sigma}\ }{\ m\ },
\end{equation}
in which $G(q)$ should satisfy \eqref{eq:DEQofG}.

In the exponent of \eqref{eq:kernel-a3}, by substituting definitions of $C_{1}(q)$ and $C_{2}(q)$ in \eqref{eq:coeffs02}, we can find the effective potential
\begin{equation}
\label{eq:Veff-a2}
V_{\mathrm{eff}}(q)=\frac{\ \hbar^{2}\ }{\ m\ }\left\{
3\{a_{3}(q)-\alpha a_{2}{'}(q)\}-2\{a_{2}(q)\}^{2}+
3a_{2}{'}(q)
\right\}
\end{equation}
in which $a_{2}(q)$ and $a_{3}(q)$ are given by \eqref{eq:coeffs01} and they are related to $h(q)$ through \eqref{eq:kleinert14}. By making use of these relations, we find the explicit form of the effective potential as
\begin{equation}
\label{eq:Veff-a3}
V_{\mathrm{eff}}(q)=\frac{\ \hbar^{2}\ }{\ 2m\ }\left[
\left\{\frac{\ 3\ }{\ 4\ }+\phi_{2}(\lambda)\right\}
\left\{\frac{\ h{''}(q)\ }{\ h{'}(q)\ }\right\}^{2}-
\left\{\frac{\ 1\ }{\ 2\ }+\phi_{3}(\lambda)\right\}
\frac{\ h{'''}(q)\ }{\ h{'}(q)\ }
\right],
\end{equation}
where we have defined
\begin{equation}
\phi_{2}(\lambda)\equiv\frac{\ 1\ }{\ 4\ }(1-2\lambda)(5+2\lambda),\quad
\phi_{3}(\lambda)\equiv\frac{\ 3\ }{\ 2\ }(1-2\lambda).
\end{equation}
Note that the ordering parameter $\alpha$ has disappeared from the coefficients by complete cancellation and further that both $\phi_{2}(\lambda)$ and $\phi_{3}(\lambda)$ vanish at the same time if we set $\lambda=1/2$.

Completing the derivation of the effective potential for the Feynman kernel, we now proceed to find the corresponding one for the fixed-energy amplitude. 
The remaining task for us is just to take the prefactor given by \eqref{eq:kleinert12} into account. The power series in \eqref{eq:kleinert13} will be exponentiated as
\begin{equation}
\label{eq:referee02}
\frac{\ h{'}(q_{j-1})\ }{\ h{'}(q_{j})\ }=\exp\left[-b_{2}(q^{(\alpha)}_{j})\Delta q_{j}-
\left(\alpha-\frac{\ 1\ }{\ 2\ }\right)\left\{
b_{3}(q^{(\alpha)}_{j})-\{b_{2}(q^{(\alpha)}_{j})\}^{2}\right\}
(\Delta q_{j})^{2}+\cdots\right].
\end{equation}
It then yields
\begin{equation}
\begin{aligned}
&\prod_{j=1}^{N}\left\{\frac{\ f(x_{j-1})\ }{\ f(x_{j})\ }\right\}^{\gamma/2}\\
=&
\exp\left[-\gamma\sum_{j=1}^{N}\left\{b_{2}(q^{(\alpha)}_{j})\Delta q_{j}+
\left(\alpha-\frac{\ 1\ }{\ 2\ }\right)\left\{
b_{3}(q^{(\alpha)}_{j})-\{b_{2}(q^{(\alpha)}_{j})\}^{2}\right\}
(\Delta q_{j})^{2}+\cdots\right\}\right].
\end{aligned}
\end{equation}

Since the left hand side of \eqref{eq:referee02} is independent of the ordering parameter $\alpha$, a comment will be useful here on the $\alpha$-dependence in the right hand side.
If we have made use of the Taylor expansion \eqref{eq:referee01} instead of \eqref{eq:kleinert13}, we will here obtain another expression given by
\begin{equation}
\label{eq:referee03}
\frac{\ h{'}(q_{j-1})\ }{\ h{'}(q_{j})\ }=
\exp\left[-b_{2}(q_{j})\Delta q_{j}
+\frac{\ 1\ }{\ 2\ }\left\{
b_{3}(q_{j})-\{b_{2}(q_{j})\}^{2}\right\}
(\Delta q_{j})^{2}+\cdots\right].
\end{equation}
The exponent of this expression can be rewritten in terms of $q^{(\alpha)}_{j}=q_{j}-\alpha\Delta q_{j}$ as
\begin{equation}
\begin{aligned}
&-b_{2}(q_{j})\Delta q_{j}-
+\frac{\ 1\ }{\ 2\ }\left\{
b_{3}(q_{j})-\{b_{2}(q_{j})\}^{2}\right\}
(\Delta q_{j})^{2}\\
=&
-\{b_{2}(q^{(\alpha)}_{j})+\alpha b_{2}{'}(q^{(\alpha)}_{j})\Delta q_{j}\}
\Delta q_{j}+\frac{\ 1\ }{\ 2\ }\left\{
b_{3}(q^{(\alpha)}_{j})-\{b_{2}(q^{(\alpha)}_{j})\}^{2}\right\}
(\Delta q_{j})^{2}+O((\Delta q_{j})^{3})\\
=&
-b_{2}(q^{(\alpha)}_{j})\Delta q_{j}-
\left(\alpha-\frac{\ 1\ }{\ 2\ }\right)\left\{
b_{3}(q^{(\alpha)}_{j})-\{b_{2}(q^{(\alpha)}_{j})\}^{2}\right\}
(\Delta q_{j})^{2}+O((\Delta q_{j})^{3}),
\end{aligned}
\end{equation}
where use has been made of the relation
\begin{equation}
	b_{2}{'}(q)=b_{3}(q)-\{b_{2}(q)\}^{2}.
\end{equation}
It will be now evident that different expressions for the ratio $h(q_{j-1})/h(q_{j})$ given by \eqref{eq:referee02} and \eqref{eq:referee03} are equivalent under a path integral provided that we discard irrelevant terms.

To make the kinetic term of the time sliced path integral be in the standard form, we make again the change of variables \eqref{eq:change01}.
Since $q^{(\alpha)}_{j}=q{'}^{(\alpha)}_{j}+O((\Delta q{'}_{j})^{2})$, $b_{n}(q^{(\alpha)}_{j})$ can be replaced by $b_{n}(q{'}^{(\alpha)}_{j})$ for $n=2,\,3$ in the above. However, since $\Delta q_{j}=\Delta q{'}_{j}-a_{2}(q{'}^{(\alpha)}_{j})(\Delta q{'}_{j})^{2}+O((\Delta q{'}_{j})^{3})$,
the first term given by $b_{2}(q^{(\alpha)}_{j})\Delta q_{j}$ produces an additional contribution to the coefficient of $(\Delta q{'}_{j})^{2}$ to yield
\begin{equation}
\begin{aligned}
&\prod_{j=1}^{N}\left\{\frac{\ f(x_{j-1})\ }{\ f(x_{j})\ }\right\}^{\gamma/2}=
\exp\left[-\gamma\sum_{j=1}^{N}\left\{b_{2}(q{'}^{(\alpha)}_{j})\Delta q{'}_{j}+
\vphantom{\frac{1}{2}}\right.\right.\\
&\left.\left.\left[\left(\alpha-\frac{\ 1\ }{\ 2\ }\right)\left\{
b_{3}(q{'}^{(\alpha)}_{j})-\{b_{2}(q{'}^{(\alpha)}_{j})\}^{2}\right\}-
b_{2}(q{'}^{(\alpha)}_{j})a_{2}(q{'}^{(\alpha)}_{j})\right]
(\Delta q{'}_{j})^{2}+\cdots\right\}
\vphantom{\sum_{j=1}^{N}}\right].
\end{aligned}
\end{equation}
This will generate, when combined together with the Jacobian factor \eqref{eq:jacobian02}, shifts on $C_{1}(q{'}^{(\alpha)}_{j})$ and $C_{2}(q{'}^{(\alpha)}_{j})$ in \eqref{eq:coeffs02} to result in
\begin{equation}
\begin{aligned}
\tilde{C}_{1}(q{'}^{(\alpha)}_{j})=&C_{1}(q{'}^{(\alpha)}_{j})+
\gamma b_{2}(q{'}^{(\alpha)}_{j})\\
\tilde{C}_{2}(q{'}^{(\alpha)}_{j})=&C_{2}(q{'}^{(\alpha)}_{j})+
(2\alpha-1)\gamma\left\{b_{3}(q{'}^{(\alpha)}_{j})-
\{b_{2}(q{'}^{(\alpha)}_{j})\}^{2}\right\}+
\gamma(1-2\lambda)^{2} \{b_{2}(q{'}^{(\alpha)}_{j})\}^{2},
\end{aligned}
\end{equation}
where the use of $2a_{2}(q{'}^{(\alpha)}_{j})=-(1-2\lambda)b_{2}(q{'}^{(\alpha)}_{j})$ has been made.
By remembering $\gamma =3(1-2\lambda)/2$, we find explicit forms of these to be given by
\begin{equation}
\label{eq:coeffs03}
\begin{aligned}
\tilde{C}_{1}(q{'}^{(\alpha)}_{j})=&
\frac{\ 1\ }{\ 2\ }(1-2\lambda)b_{2}(q{'}^{(\alpha)}_{j})\\
\tilde{C}_{2}(q{'}^{(\alpha)}_{j})=&
\left\{\frac{\ 3\ }{\ 4\ }+(1-\alpha)(1-2\lambda)+
\frac{\ 1\ }{\ 4\ }(1-2\lambda)^{2}\right\}
\{b_{2}(q^{(\alpha)}_{j})\}^{2}-\\
&\left\{\frac{\ 1\ }{\ 2\ }+(1-\alpha)(1-2\lambda)\right\}
b_{3}(q^{(\alpha)}_{j}).
\end{aligned}
\end{equation}
In this way, the $f$-factor $\{f(x_{0})/f(x_{N})\}^{\gamma/2}$, combined with the Jacobian factor, contributes
\begin{equation}
\label{eq:jacobian03}
\exp\left[
-\sum_{j=1}^{N}\left\{\tilde{C}_{1}(q{'}^{(\alpha)}_{j})\Delta q{'}_{j}+
\frac{\ 1\ }{\ 2\ }\tilde{C}_{2}(q{'}^{(\alpha)}_{j})
(\Delta q{'}_{j})^{2}\right\}
\right]
\end{equation}
to the time sliced path integral. It will be useful to observe here that the combination $\tilde{C}_{2}(q)+2(1-\alpha)\tilde{C}_{1}{'}(q)$ is independent of $\alpha$ since $b_{2}{'}(q)=b_{3}(q)-\{b_{2}(q)\}^{2}$.

We follow again the same procedure given in the appendix here to obtain
\begin{equation}
\label{eq:Veff-g0}
\begin{aligned}
V_{\mathrm{eff}}(q)=&\frac{\ \hbar^{2}\ }{\ 2m\ }\left\{
\{\tilde{C}_{1}(q)\}^{2}+\tilde{C}_{2}(q)+
2(1-\alpha)\tilde{C}_{1}{'}(q)
\right\}\\
=&
\frac{\ \hbar^{2}\ }{\ 2m\ }\left[
\left\{\frac{\ 3\ }{\ 4\ }+\frac{\ 1\ }{\ 2\ }(1-2\lambda)^{2}\right\}
\left\{\frac{\ h{''}(q)\ }{\ h{'}(q)\ }\right\}^{2}-
\frac{\ 1\ }{\ 2\ }\frac{\ h{'''}(q)\ }{\ h{'}(q)\ }
\right]
\end{aligned}
\end{equation}
as the effective potential for the time sliced path integral of the fixed-energy amplitude. In the process of calculating the effective potential, the ordering parameter $\alpha$ has been cancelled  out completely to ensure the ordering independence of the time sliced path integral.
On the other hand, the effective potential still depends on the splitting parameter $\lambda$ even in the time sliced path integral for the fixed-energy amplitude, though in a slightly different way when compared to that of the Feynman kernel. If we set $\lambda=1/2$, however, they coincide with each other and also with the one given in chapter 14 of Ref.\onlinecite{KleinertBook}.

The explicit form of the time sliced path integral for the Feynman kernel is now found to be given by
\begin{equation}
\label{eq:pi-kernel}
\begin{aligned}
&K_{E}(x,x{'};\sigma)=\frac{\ 1\ }{\ \sqrt{f_{l}(x)f_{r}(x{'})\,}\ }
\left(\frac{\ m\ }{\ 2\pi\hbar\epsilon_{\sigma}\ }\right)^{N/2}
\int_{-\infty}^{\infty}\!\prod_{i=1}^{N-1}d\xi_{i}\\
&\times
\exp\left[-\sum_{j=1}^{N}\left\{\frac{\ m\ }{\ 2\hbar\epsilon_{\sigma}\ }
(\Delta \xi_{j})^{2}+
\frac{\ \epsilon_{\sigma}\ }{\ \hbar\ }
\{h{'}(\xi^{(\alpha)}_{j})\}^{2}\{V(h(\xi^{(\alpha)}_{j}))-E\}\right\}\right]\\
&\times
\exp\left[
-\frac{\ \hbar\epsilon_{\sigma}\ }{\ 2m\ }\sum_{j=1}^{N}\left[
\left\{\frac{\ 3\ }{\ 4\ }+\phi_{2}(\lambda)\right\}
\left\{\frac{\ h{''}(\xi^{(\alpha)}_{j})\ }{\ h{'}(\xi^{(\alpha)}_{j})\ }\right\}^{2}-
\left\{\frac{\ 1\ }{\ 2\ }+\phi_{3}(\lambda)\right\}
\frac{\ h{'''}(\xi^{(\alpha)}_{j})\ }{\ h{'}(\xi^{(\alpha)}_{j})\ }\right]\right]
\end{aligned}
\end{equation}
and the corresponding one for the fixed-energy amplitude reads
\begin{equation}
\label{eq:pi-green}
\begin{aligned}
&G(x,x{'};E)=\{f(x)f(x{'})\}^{1/4}\int_{0}^{\infty}\frac{\ d\sigma\ }{\ \hbar\ }
\left(\frac{\ m\ }{\ 2\pi\hbar\epsilon_{\sigma}\ }\right)^{N/2}
\int_{-\infty}^{\infty}\!\prod_{i=1}^{N-1}d\xi_{i}\\\
&\times
\exp\left[-\sum_{j=1}^{N}\left\{\frac{\ m\ }{\ 2\hbar\epsilon_{\sigma}\ }
(\Delta \xi_{j})^{2}+
\frac{\ \epsilon_{\sigma}\ }{\ \hbar\ }
\{h{'}(\xi^{(\alpha)}_{j})\}^{2}\{V(h(\xi^{(\alpha)}_{j}))-E\}\right\}\right]\\
&\times
\exp\left[
-\frac{\ \hbar\epsilon_{\sigma}\ }{\ 2m\ }\sum_{j=1}^{N}\left[
\left\{\frac{\ 3\ }{\ 4\ }+\frac{\ 1\ }{\ 2\ }(1-2\lambda)^{2}\right\}
\left\{\frac{\ h{''}(\xi^{(\alpha)}_{j})\ }{\ h{'}(\xi^{(\alpha)}_{j})\ }\right\}^{2}-
\frac{\ 1\ }{\ 2\ }
\frac{\ h{'''}(\xi^{(\alpha)}_{j})\ }{\ h{'}(\xi^{(\alpha)}_{j})\ }\right]\right].
\end{aligned}
\end{equation}
In view of the definition of $\xi_{j}$, given by \eqref{eq:q'2xi},
a change in the ordering parameter $\alpha$ will cause a corresponding alternation in $\xi_{j}$. The change in $\xi_{j}$ is, however, just a change or a renaming of the variable of integration into another one. Therefore the possible change in path integrals above under a shift of ordering parameter $\alpha\mapsto\alpha{'}$ is restricted to the transformation from $\xi^{(\alpha)}_{j}$ to $\xi^{(\alpha{'})}_{j}$ in the local potential term as well as the effective potential. As will be evident from the existence of $\epsilon_{\sigma}$, $V_{\mathrm{eff}}(\xi^{(\alpha{'})}_{j})$ is equivalent to $V_{\mathrm{eff}}(\xi^{(\alpha)}_{j})$ under path integrals above because they differ only in irrelevant terms. From these consideration, we are convinced that these path integrals are independent of the ordering parameter $\alpha$.

In the derivation of time sliced path integrals above, we have experienced two kinds of change of variables in addition to the original one from $x_{j}$'s to $q_{j}$'s defined by $x_{j}=h(q_{j})$. The first step of the additional change of variables was achieved by introducing $q{'}_{j}$ to be different from $q_{j}$ by a power series of $\Delta q_{k}$($k=1,\,2,\,\dots,\,j$). This step enables us to express $\Delta q_{j}R(q^{(\alpha)}_{j},\Delta q_{j})$ as $\Delta q{'}_{j}$ but, from the Jacobian, first and second order terms of $\Delta q{'}_{j}$ can be generated in the exponent of time sliced path integrals. To absorb these terms into the kinetic term, we have needed to introduce $\xi_{j}$'s by \eqref{eq:q'2xi} as the second step of the additional change of variables. It is the Jacobian of this second step that mainly determines the effective potential of the time sliced path integral. At this stage, we have to emphasize that we could not find the precise form of the effective potential without formulating the path integral with finite, though infinitesimally small, pseudotime intervals because $\Delta q_{j}R(q^{(\alpha)}_{j},\Delta q_{j})/\epsilon_{\sigma}$ simply yields $dq/ds$($s=j\epsilon_{\sigma}$) in the naive continuum limit. The transformation from $x_{j}$'s to $\xi_{j}$'s utilized above has, therefore, no classical counterpart and hence cannot be a point canonical transformation\cite{GirottiSimoes,FukutakaKashiwa,KashiwaBook}. 

Before closing this section, let us observe the mechanism that causes the cancellation in the possibly $\alpha$-dependent terms in the effective potential.
We see from \eqref{eq:coeffs01} that
\begin{equation}
\label{eq:cancellation01}
a_{3}(q)=\frac{\ 1\ }{\ 2\ }\left\{
\lambda^{2}\{b_{2}(q)\}^{2}-
\left(\lambda-\frac{\ 1\ }{\ 3\ }\right)b_{3}(q)\right\}-
\frac{\ \alpha\ }{\ 2\ }(1-2\lambda)\left\{
b_{3}(q)-\{b_{2}(q)\}^{2}\right\}
\end{equation}
and
\begin{equation}
\label{eq:cancellation02}
a_{2}{'}(q)=-\frac{\ 1\ }{\ 2\ }(1-2\lambda)
\left\{b_{3}(q)-\{b_{2}(q)\}^{2}\right\}.
\end{equation}
Therefore, for the system we are dealing with, there happens to hold
\begin{equation}
\label{eq:cancellation03}
a_{3}(q)-\alpha a_{2}{'}(q)=\frac{\ 1\ }{\ 2\ }\left\{
\lambda^{2}\{b_{2}(q)\}^{2}-
\left(\lambda-\frac{\ 1\ }{\ 3\ }\right)b_{3}(q)\right\}
\end{equation}
to make this combination to be independent of $\alpha$.
As has been already pointed out, a similar relation can be seen from \eqref{eq:coeffs03} for the combination
$\tilde{C}_{2}(q)+2(1-\alpha)\tilde{C}_{1}{'}(q)$ in the effective potential \eqref{eq:Veff-g0}. These are the reason for the $\alpha$-independence of time sliced path integrals we have considered above.

\section{Application to the radial path integral}

Completing our main task of this article, we here try to find its application to some wellknown models in this section. An example will be the radial path integral of the hydrogen atom in three dimensions\cite{Inomata}. The model is defined by
\begin{equation}
\label{eq:inomata01}
\begin{aligned}
Q_{l}(r,r{'};\tau)=&\frac{\ 1\ }{\ rr{'}\ }
\left(\frac{\ m\ }{\ 2\pi\hbar\epsilon\ }\right)^{N/2}
\int_{0}^{\infty}\!\prod_{i=1}^{N-1}dr_{i}\\
&\times
\exp\left[-\sum_{j=1}^{N}\left\{\frac{\ m\ }{\ 2\hbar\epsilon\ }
(\Delta r_{j})^{2}+\frac{\ l(l+1)\hbar\epsilon\ }{\ 2mr_{j}r_{j-1}\ }-
\frac{\ \epsilon\ }{\ \hbar\ }
\left(\frac{\ \kappa\ }{\ \sqrt{r_{j}r_{j-1}\,}\ }+E\right)\right\}
\right]
\end{aligned}
\end{equation}
where we have employed the geometric mean for the $1/r$ potential.
The corresponding fixed-energy amplitude is given by
\begin{equation}
\label{eq:inomata02}
G_{l}(r,r{'};E)=\frac{\ 1\ }{\ \hbar\ }\int_{0}^{\infty}\!d\tau
Q_{l}(r,r{'};\tau).
\end{equation}
The time sliced path integral for this Green's function will be formulated by utilizing the one for $Q_{l}(r,r{'};\tau)$ given above.
According to Inomata, making a change of variables from $r_{j}$ to $\rho_{j}=r_{j}^{1/2}$ as well as the introduction of a new time interval $\sigma_{j}=\epsilon/(4\sqrt{r_{j}r_{j-1}\,})$ seems to convert the above time sliced path integral into the one for the radial oscillator in three dimensions.
However the path collapse\cite{KleinertBook} must be avoided before discussing the detail of a time sliced path integral for singular potentials. We, therefore, consider here the time sliced path integral for a new Hamiltonian
\begin{equation}
\label{eq:inomata03}
H_{E}=f_{l}(r)\left[
-\frac{\ \hbar^{2}\ }{\ 2m\ }\frac{\ d^{2}\ \ }{\ dr^{2}\ }
+\frac{\ \hbar^{2}\ }{\ 2m\ }\frac{\ \mu^{2}-1/4\ }{\ r^{2}\ }
+V(r)-E\right]f_{r}(r)
\end{equation}
where $\mu=l+D/2-1$($l=0,\,1,\,2,\,\dots$) for a system with potential $V(r)$ in $D$-dimensional space.
In the following, we make a special choice for the regulating function $f(r)$ by setting $f_{l}(r)=f_{r}(r)=r$(i.e. $\lambda=1/2$) so that $f(r)=r^{2}$.
The corresponding new variable $q$ is defined by $r=h(q)$ and $f(r)=\{h{'}(q)\}^{2}$. These are solved by setting $h(q)=ae^{q}$ where $a$ is a positive constant that carries the dimension of length.

In order to formulate path integrals for systems on a half line, it will be useful to work with the ``radial plane wave"\cite{Fujikawa2008} given by
\begin{equation}
\label{eq:inomata04}
\langle r\vert p\rangle=\frac{\ 1\ }{\ \sqrt{2\pi\hbar\,}\ }
e^{ipr/\hbar}.
\end{equation}
For this wave function the completeness or resolution of unity holds:
\begin{equation}
\label{eq:inomata05}
\int_{-\infty}^{\infty}\!\!dp\,
\langle r\vert p\rangle\langle p\vert r{'}\rangle=\delta(r-r{'})
\end{equation}
though the orthogonality relation is not satisfied.
By making use of the completeness \eqref{eq:inomata05}, we find
\begin{equation}
\label{eq:inomata06}
\begin{aligned}
&\left\langle r_{j}\left\vert
\left(1-\frac{\ \epsilon_{\sigma}\ }{\ \hbar\ }H_{E}\right)
\right\vert r_{j-1}\right\rangle=
\int\!\frac{\ dp\ }{\ 2\pi\hbar\ }
\exp\left[\frac{\ i \ }{\ \hbar\ }p\Delta r_{j}-
\right.\\
&\hphantom{\left(1-\frac{\ \epsilon_{\sigma}\ }{\ \hbar\ }H_{E}\right)}
\left.\frac{\ \epsilon_{\sigma} \ }{\ \hbar\ }\left\{
\frac{\ 1\ }{\ 2m\ }r_{j}r_{j-1}p^{2}+
\frac{\ \hbar^{2}\ }{\ 2m\ }(\mu^{2}-1/4)+
r_{j}r_{j-1}\{V(\bar{r}_{j})-E\}
\right\}\right],
\end{aligned}
\end{equation}
where $\Delta r_{j}=r_{j}-r_{j-1}$ and $\bar{r}_{j}=\sqrt{r_{j}r_{j-1}\,}$. After carrying out the integration with respect to $p$, we obtain
\begin{equation}
\begin{aligned}
\label{eq:inomata07}
\left\langle r_{j}\left\vert
\left(1-\frac{\ \epsilon_{\sigma}\ }{\ \hbar\ }H_{E}\right)
\right\vert r_{j-1}\right\rangle=&
\sqrt{\frac{\ m\ }{\ 2\pi\hbar\epsilon_{\sigma}r_{j}r_{j-1}\ }\,}
\exp\left[-\frac{\ m\ }{\ 2\hbar\epsilon_{\sigma}\ }
\left(\frac{\ \Delta r_{j}\ }{\ \sqrt{r_{j}r_{j-1}\,}\ }\right)^{2}-
\right.\\
&\left.
\frac{\ \hbar\epsilon_{\sigma}\ }{\ 2m\ }(\mu^{2}-1/4)-
\frac{\ \epsilon_{\sigma}\ }{\ \hbar\ }r_{j}r_{j-1}
\{V(\bar{r}_{j})-E\}\right].
\end{aligned}
\end{equation}

Under the change of variables given by $r_{j}=h(q_{j})$, the kinetic term of the short time kernel is rewritten as
\begin{equation}
\label{eq:inomata08}
\left(\frac{\ \Delta r_{j}\ }{\ \sqrt{r_{j}r_{j-1}\,}\ }\right)^{2}=
\{\Delta q_{j}R(\bar{q}_{j},\Delta q_{j})\}^{2},
\end{equation}
where $\Delta q_{j}=q_{j}-q_{j-1}$, $\bar{q}_{j}=(q_{j}+q_{j-1})/2$ and $R(\bar{q}_{j},\Delta q_{j})$ being given by
\begin{equation}
\label{eq:inomata09}
R(\bar{q}_{j},\Delta q_{j})=\frac{\ 2\ }{\ \Delta q_{j}\ }
\sinh(\Delta q_{j}/2)=
1+\frac{\ 1\ }{\ 24\ }(\Delta q_{j})^{2}+O((\Delta q_{j})^{4}).
\end{equation}
In view of this series expansion, we find that $R(\bar{q}_{j},\Delta q_{j})$ is actually independent of $\bar{q}_{j}$ and further that $a_{2}(\bar{q}_{j})=0$ and $a_{3}(\bar{q}_{j})=1/24$ for this system. Therefore the effective potential is given by
\begin{equation}
\label{eq:inomata10}
V_{\mathrm{eff}}(q)=\frac{\ \hbar^{2}\ }{\ 8m\ }
\end{equation}
which will remove $1/4$ from $\mu^{2}-1/4$ by cancellation in the exponent of \eqref{eq:inomata07}.
By taking these into account, we obtain a time sliced path integral for the Feynman kernel of the Hamiltonian \eqref{eq:inomata03}
\begin{equation}
\label{eq:inomata11}
\begin{aligned}
K_{E}(r,r{'};\sigma)=&\frac{\ 1\ }{\ \sqrt{rr{'}\,}\ }
\left(\frac{\ m\ }{\ 2\pi\hbar\epsilon_{\sigma}\ }\right)^{N/2}
\int_{-\infty}^{\infty}\!\prod_{i=1}^{N-1}dq_{i}\,
\exp\left[-\sum_{j=1}^{N}\frac{\ m\ }{\ 2\hbar\epsilon_{\sigma}\ }
(\Delta q_{j})^{2}\right]\\
&\times
\exp\left[
-\frac{\ \epsilon_{\sigma}\ }{\ \hbar\ }\sum_{j=1}^{N}\left\{
\frac{\ \hbar^{2}\mu^{2}\ }{\ 2m\ }+
a^{2}e^{2\bar{q}_{j}}\{V(ae^{\bar{q}_{j}})-E\}\right\}\right].
\end{aligned}
\end{equation}

Another formulation of the radial path integral above is possible if we make use of the completeness of eigenfunctions of the free Hamiltonian
\begin{equation}
\label{eq:bessel01}
H_{0}=-\frac{\ \hbar^{2}\ }{\ 2m\ }\frac{\ d^{2}\ \ }{\ dr^{2}\ }
+\frac{\ \hbar^{2}\ }{\ 2m\ }\frac{\ \mu^{2}-1/4\ }{\ r^{2}\ }
\end{equation}
whose eigenfunction being given, in terms of the Bessel function, by $\psi_{k}(r)=\sqrt{kr\,}J_{\mu}(kr)$($k>0$). The eigenfunction $\psi_{k}(r)=\langle r\vert k\rangle$ obeys
\begin{equation}
\label{eq:bessel02}
H_{0}\psi_{k}(r)=\frac{\ \hbar^{2}k^{2}\ }{\ 2m\ }\psi_{k}(r).
\end{equation}
The completeness
\begin{equation}
\label{eq:bessel03}
\int_{0}^{\infty}\!\!dk\langle r\vert k\rangle\langle k\vert r{'}\rangle=
\delta(r-r{'})
\end{equation}
holds and the orthogonality relation
\begin{equation}
\label{eq:bessel04}
\int_{0}^{\infty}\!\!dr\langle k\vert r\rangle\langle r\vert k{'}\rangle=
\delta(k-k{'})
\end{equation}
is fulfilled. These are the consequence of the formula($\beta>0$)
\begin{equation}
\label{eq:bessel05}
\int_{0}^{\infty}\!\!e^{-\beta x^{2}}xJ_{\mu}(px)J_{\mu}(qx)\,dx=
\frac{\ 1\ }{\ 2\beta\ }e^{-(p^{2}+q^{2})/(4\beta)}
I_{\mu}\left(\frac{\ pq\ }{\ 2\beta\ }\right),
\end{equation}
where $I_{\mu}(z)$ denotes the modified Bessel function.
By making use of the completeness \eqref{eq:bessel03} and by setting $f_{l}(r)=f_{r}(r)=r$ again, we obtain
\begin{equation}
\label{eq:bessel06}
\begin{aligned}
\left\langle r_{j}\left\vert
\left(1-\frac{\ \epsilon_{\sigma}\ }{\ \hbar\ }H_{E}\right)
\right\vert r_{j-1}\right\rangle=&
\int_{0}^{\infty}\!\!kdk\sqrt{r_{j}r_{j-1}\,}
J_{\mu}(kr_{j})J_{\mu}(kr_{j-1})\times\\
&
\exp\left[-\frac{\ \epsilon_{\sigma}\ }{\ \hbar\ }r_{j}r_{j-1}\left\{
\frac{\ \hbar^{2}k^{2}\ }{\ 2m\ }+V(\bar{r}_{j})-E\right\}\right].
\end{aligned}
\end{equation}
Then, by carrying out the integration with respect to $k$, we find
\begin{equation}
\label{eq:bessel07}
\begin{aligned}
&\left\langle r_{j}\left\vert
\left(1-\frac{\ \epsilon_{\sigma}\ }{\ \hbar\ }H_{E}\right)
\right\vert r_{j-1}\right\rangle\\
=&
\frac{\ m\ }{\ \hbar\epsilon_{\sigma}\sqrt{r_{j}r_{j-1}\,}\ }
\exp\left[-\frac{\ m\ }{\ 2\hbar\epsilon_{\sigma}\ }
\left(\frac{\ r_{j}\ }{\ r_{j-1}\ }+\frac{\ r_{j-1}\ }{\ r_{j}\ }\right)
-\frac{\ \epsilon_{\sigma}\ }{\ \hbar\ }r_{j}r_{j-1}\{V(\bar{r}_{j})-E\}\right]
I_{\mu}\left(\frac{\ m\ }{\ \hbar\epsilon_{\sigma}\ }\right).
\end{aligned}
\end{equation}
Thanks to the symmetric setting for the regulating function, we have obtained $r$-independent expression in the argument of the modified Bessel function.

We now make a change of variables from $r_{j}$ to $q_{j}$ by setting $r_{j}=ae^{q_{j}}$($a>0$) to obtain a time sliced path integral
\begin{equation}
\label{eq:bessel08}
\begin{aligned}
&K_{E}(r,r{'};\sigma)=\frac{\ 1\ }{\ \sqrt{rr{'}\,}\ }
\left\{\frac{\ m\ }{\ \hbar\epsilon_{\sigma}\ }e^{-m/(\hbar\epsilon_{\sigma})}
I_{\mu}\left(\frac{\ m\ }{\ \hbar\epsilon_{\sigma}\ }\right)\right\}^{N}\times\\
&\int_{-\infty}^{\infty}\!\prod_{i=1}^{N-1}dq_{i}
\exp\left[-\frac{\ m\ }{\ 2\hbar\epsilon_{\sigma}\ }\sum_{j=1}^{N}
\left\{\Delta q_{j}R(\bar{q}_{j},\Delta q_{j})\right\}^{2}-
\frac{\ \epsilon_{\sigma}\ }{\ \hbar\ }\sum_{j=1}^{N}a^{2}e^{2\bar{q}_{j}}
\{V(ae^{\bar{q}_{j}})-E\}\right],
\end{aligned}
\end{equation}
where $R(\bar{q}_{j},\Delta q_{j})$ is given by 
$2\sinh(\Delta q_{j}/2)/\Delta q_{j}$ again to generate the same effective potential as before. In the limit $\epsilon_{\sigma}\to0$ we can set
\begin{equation}
\label{eq:bessel09}
\frac{\ m\ }{\ \hbar\epsilon_{\sigma}\ }e^{-m/(\hbar\epsilon_{\sigma})}
I_{\mu}\left(\frac{\ m\ }{\ \hbar\epsilon_{\sigma}\ }\right)=
\sqrt{\frac{\ m\ }{\ 2\pi\hbar\epsilon_{\sigma}\ }\,}
\exp\left\{-\frac{\ \hbar\epsilon_{\sigma}\ }{\ 2m\ }(\mu^{2}-1/4)\right\}
\end{equation}
to obtain the time sliced path integral
\begin{equation}
\label{eq:bessel10}
\begin{aligned}
K_{E}(r,r{'};\sigma)=&\frac{\ 1\ }{\ \sqrt{rr{'}\,}\ }
\left(\frac{\ m\ }{\ 2\pi\hbar\epsilon_{\sigma}\ }\right)^{N/2}
\int_{-\infty}^{\infty}\!\prod_{i=1}^{N-1}dq_{i}\,
\exp\left[-\sum_{j=1}^{N}\frac{\ m\ }{\ 2\hbar\epsilon_{\sigma}\ }
(\Delta q_{j})^{2}\right]\\
&\times
\exp\left[
-\frac{\ \epsilon_{\sigma}\ }{\ \hbar\ }\sum_{j=1}^{N}\left\{
\frac{\ \hbar^{2}\mu^{2}\ }{\ 2m\ }+
a^{2}e^{2\bar{q}_{j}}\{V(ae^{\bar{q}_{j}})-E\}\right\}\right]
\end{aligned}
\end{equation}
which is identical to \eqref{eq:inomata11}.
We have thus confirmed that both the radial plane wave and the eigenfunction of the free Hamiltonian \eqref{eq:bessel01} yield the same time sliced path integral for the Feynman kernel. Note that we have employed the mid-point prescription $V(r)\mapsto V(\bar{r}_{j})$ in the above derivation but the factor $e^{2\bar{q}_{j}}$ is the consequence of the symmetric setting $f_{l}(r)=f_{r}(r)=r$ for the regulating function. If we have set
 $f_{l}(r)=r^{2\lambda}$ and $f_{r}(r)=r^{2(1-\lambda)}$ instead,
 we could have obtained $e^{2\bar{q}_{j}+(2\lambda-1)\Delta q_{j}}$ in front of $V(ae^{\bar{q}_{j}})-E$. Due to the existence of $\epsilon_{\sigma}$, we can discard $\Delta q_{j}$ in the above and replace $e^{2\bar{q}_{j}+(2\lambda-1)\Delta q_{j}}$ by $e^{2\bar{q}_{j}}$. However the effective potential becomes $\lambda$ dependent to result in a different form of the time sliced path integral.

Let us now consider special cases: (i) the radial oscillator and (ii) the Coulomb potential by setting $V(r)=m\omega^{2}r^{2}/2$ and $V(r)=-\kappa/r$, respectively. For the radial oscillator, the time sliced path integral is expressed as
\begin{equation}
\label{eq:oscillator01}
\begin{aligned}
K_{E}^{(\mathrm O)}(r,r{'};\sigma)=&\frac{\ 1\ }{\ \sqrt{rr{'}\,}\ }
\left(\frac{\ m\ }{\ 2\pi\hbar\epsilon_{\sigma}\ }\right)^{N/2}
\int_{-\infty}^{\infty}\!\prod_{i=1}^{N-1}dq_{i}\,
\exp\left[-\sum_{j=1}^{N}\frac{\ m\ }{\ 2\hbar\epsilon_{\sigma}\ }
(\Delta q_{j})^{2}\right]\\
&\times
\exp\left[
-\frac{\ \epsilon_{\sigma}\ }{\ \hbar\ }\sum_{j=1}^{N}\left\{
\frac{\ \hbar^{2}\mu^{2}\ }{\ 2m\ }+
\frac{\ 1\ }{\ 2\ }m\omega^{2}a^{4}e^{4\bar{q}_{j}}
-Ea^{2}e^{2\bar{q}_{j}}\right\}\right]
\end{aligned}
\end{equation}
while the corresponding one for the Coulomb potential being given by
\begin{equation}
\label{eq:Coulomb01}
\begin{aligned}
K_{E_{\mathrm C}}^{(\mathrm C)}(r,r{'};\sigma)=&\frac{\ 1\ }{\ \sqrt{rr{'}\,}\ }
\left(\frac{\ m_{\mathrm C}\ }{\ 2\pi\hbar\epsilon_{\sigma}\ }\right)^{N/2}
\int_{-\infty}^{\infty}\!\prod_{i=1}^{N-1}dq_{i}\,
\exp\left[-\sum_{j=1}^{N}\frac{\ m_{\mathrm C}\ }{\ 2\hbar\epsilon_{\sigma}\ }
(\Delta q_{j})^{2}\right]\\
&\times
\exp\left[
-\frac{\ \epsilon_{\sigma}\ }{\ \hbar\ }\sum_{j=1}^{N}\left\{
\frac{\ \hbar^{2}\mu_{\mathrm C}^{2}\ }{\ 2m_{\mathrm C}\ }-
\kappa ae^{\bar{q}_{j}}
-E_{\mathrm C}a^{2}e^{2\bar{q}_{j}}\right\}\right].
\end{aligned}
\end{equation}
For the radial oscillator, we set
\begin{equation}
\label{eq:oscillator02}
a=\sqrt{\frac{\ \hbar\ }{\ m\omega\ }\,},\
E=\nu\hbar\omega
\end{equation}
and scale variables $q_{j}\to q_{j}/2$ to find
\begin{equation}
\label{eq:oscillator03}
\begin{aligned}
K_{E}^{(\mathrm O)}(r,r{'};\sigma)=&\frac{\ 2\ }{\ \sqrt{rr{'}\,}\ }
\left(\frac{\ m\ }{\ 8\pi\hbar\epsilon_{\sigma}\ }\right)^{N/2}
\int_{-\infty}^{\infty}\!\prod_{i=1}^{N-1}dq_{i}\,
\exp\left[-\sum_{j=1}^{N}\frac{\ m\ }{\ 8\hbar\epsilon_{\sigma}\ }
(\Delta q_{j})^{2}\right]\\
&\times
\exp\left[
-\frac{\ \hbar\epsilon_{\sigma}\ }{\ 2m\ }\sum_{j=1}^{N}\left(
\mu^{2}+
e^{2\bar{q}_{j}}
-2\nu e^{\bar{q}_{j}}\right)\right].
\end{aligned}
\end{equation}
In the same way, we set
\begin{equation}
\label{eq:Coulomb02}
a=\frac{\ \nu_{\mathrm C}\ }{\ 2\ }a_{\mathrm B},\
E_{\mathrm C}=-\frac{\ \kappa\ }{\ 2\nu_{\mathrm C}^{2}a_{\mathrm B}\ },\quad
a_{\mathrm B}=\frac{\ \hbar^{2}\ }{\ m_{\mathrm C}\kappa\ }
\end{equation}
for the radial Coulomb system to obtain
\begin{equation}
\label{eq:Coulomb03}
\begin{aligned}
K_{E_{\mathrm C}}^{(\mathrm C)}(r,r{'};\sigma)=&\frac{\ 1\ }{\ \sqrt{rr{'}\,}\ }
\left(\frac{\ m_{\mathrm C}\ }{\ 2\pi\hbar\epsilon_{\sigma}\ }\right)^{N/2}
\int_{-\infty}^{\infty}\!\prod_{i=1}^{N-1}dq_{i}\,
\exp\left[-\sum_{j=1}^{N}\frac{\ m_{\mathrm C}\ }{\ 2\hbar\epsilon_{\sigma}\ }
(\Delta q_{j})^{2}\right]\\
&\times
\exp\left[
-\frac{\ \hbar\epsilon_{\sigma}\ }{\ 2m_{\mathrm C}\ }\sum_{j=1}^{N}\left(
\mu_{\mathrm C}^{2}-
\nu_{\mathrm C}e^{\bar{q}_{j}}+
\frac{\ 1\ }{\ 4\ }e^{2\bar{q}_{j}}\right)\right].
\end{aligned}
\end{equation}
If we set in \eqref{eq:oscillator03}
\begin{equation}
\label{eq:C2O01}
\frac{\ 1\ }{\ 4\ }m=m_{\mathrm C},\
\frac{\ 1\ }{\ 2\ }\mu=\mu_{\mathrm C},\
\frac{\ 1\ }{\ 2\ }\nu=\nu_{\mathrm C},\
\end{equation}
we find
\begin{equation}
\label{eq:C2O02}
K_{E_{\mathrm C}}^{(\mathrm C)}(r,r{'};\tau)=
\frac{\ 1\ }{\ 2\sqrt{\rho\rho{'}\,}\ }
K_{E}^{(\mathrm O)}(\rho,\rho{'};\tau),\
\rho=\sqrt{r\,},\
\rho{'}=\sqrt{r{'}\,}.
\end{equation}

The time sliced path integral for the fixed-energy amplitude that correspond to \eqref{eq:inomata11} or \eqref{eq:bessel10} will be formulated without difficulties. For the radial oscillator, we find
\begin{equation}
\label{eq:oscillator12}
\begin{aligned}
G^{(\mathrm{O})}(r,r{'};E)=&2\sqrt{rr{'}\,}
\int_{0}^{\infty}\!\frac{\ d\sigma\ }{\ \hbar\ }
\left(\frac{\ m\ }{\ 8\pi\hbar\epsilon_{\sigma}\ }\right)^{N/2}
\int_{-\infty}^{\infty}\!\prod_{i=1}^{N-1}dq_{i}\,
\exp\left[-\sum_{j=1}^{N}\frac{\ m\ }{\ 8\hbar\epsilon_{\sigma}\ }
(\Delta q_{j})^{2}\right]\\
&\times
\exp\left[
-\frac{\ \hbar\epsilon_{\sigma}\ }{\ 2m\ }\sum_{j=1}^{N}\left(
\mu^{2}+
e^{2\bar{q}_{j}}
-2\nu e^{\bar{q}_{j}}\right)\right]
\end{aligned}
\end{equation}
and for the Coulomb potential we obtain
\begin{equation}
\label{eq:Coulomb12}
\begin{aligned}
G^{(\mathrm{C})}(r,r{'};E_{\mathrm C})=&\sqrt{rr{'}\,}
\int_{0}^{\infty}\!\frac{\ d\sigma\ }{\ \hbar\ }
\left(\frac{\ m_{\mathrm C}\ }{\ 2\pi\hbar\epsilon_{\sigma}\ }\right)^{N/2}
\int_{-\infty}^{\infty}\!\prod_{i=1}^{N-1}dq_{i}\,
\exp\left[-\sum_{j=1}^{N}\frac{\ m_{\mathrm C}\ }{\ 2\hbar\epsilon_{\sigma}\ }
(\Delta q_{j})^{2}\right]\\
&\times
\exp\left[
-\frac{\ \hbar\epsilon_{\sigma}\ }{\ 2m_{\mathrm C}\ }\sum_{j=1}^{N}\left(
\mu_{\mathrm C}^{2}-
\nu_{\mathrm C}e^{\bar{q}_{j}}+
\frac{\ 1\ }{\ 4\ }e^{2\bar{q}_{j}}\right)\right].
\end{aligned}
\end{equation}
By setting the same relations among parameters above, we observe
\begin{equation}
\label{eq:C2O03}
G^{(\mathrm{C})}(r,r{'};E_{\mathrm C})=
\frac{\ 1\ }{\ 2\ }\sqrt{\rho\rho{'}\,}
G^{(\mathrm{O})}(\rho,\rho{'};E),\
\rho=\sqrt{r\,},\
\rho{'}=\sqrt{r{'}\,}.
\end{equation}
We have thus confirmed the equivalence of radial path integrals for the Coulomb potential and the isotropic oscillator.

From the relation
\begin{equation}
\label{eq:dimen01}
\mu_{\mathrm C}=\frac{\ 1\ }{\ 2\ }\mu,
\end{equation}
we find for the Coulomb path integral in three dimensions
\begin{equation}
\label{eq:dimen02}
2l_{\mathrm C}+1=l+\frac{\ D\ }{\ 2\ }-1.
\end{equation}
If we set $D=3$ in the right hand side above, we obtain\cite{Inomata}
\begin{equation}
\label{eq:dimen03}
l=2l_{\mathrm C}+\frac{\ 1\ }{\ 2\ }.
\end{equation}
Since the radial Hamiltonian is defined upon the angular decomposition, the angular momentum $l$ of the oscillator system must be an integer. Therefore the relation \eqref{eq:dimen03} is not admissible to determine the angular momentum for a radial oscillator in three dimensions. Going back to \eqref{eq:dimen01} and writing the dimension of the Coulomb system as $D_{\mathrm C}$, we get
\begin{equation}
\label{eq:dimen04}
l=2l_{\mathrm C}+D_{\mathrm C}-\frac{\ D\ }{\ 2\ }-1
\end{equation}
and find that $D$ must be an even integer satisfying $D\le2(D_{\mathrm C}-1)$.
For three-dimensional Coulomb system, possible values of the dimension of the harmonic oscillator are $D=2$ or $D=4$. By choosing $D=4$, we get $l=2l_{\mathrm C}$ for $D_{\mathrm C}=3$. This will be interpreted as the relation of angular momenta between the four-dimensional oscillator and the hydrogen atom in the path integral solution of the hydrogen atom by Duru and Kleinert. To embed the path integral of the three-dimensional Coulomb system as a whole, we may need a four-dimensional space for the path integral of the corresponding oscillator system. Nevertheless we can deal with the radial path integral itself as an independent issue. We may consider the choice $D=2$ for $D_{\mathrm C}=3$, for which we have $l=2l_{\mathrm C}+1$, as the path integral counterpart of Schwinger's trick\cite{SchwingerQM} in this sense. For two-dimensional Coulomb system($D_{\mathrm C}=2$), $D=2$ is unique as the dimension of the corresponding oscillator system. In this respect, the Coulomb system in one dimension is quite exceptional. We will not be able to find a corresponding oscillator path integral since we get $D=0$ for this case. An exact solution\cite{Sakoda} is, however, obtained even for this system by making use of the Duru-Kleinert formalism.

\section{Conclusion}

In this paper we have formulated time sliced path integrals in the DK formalism without depending on the use of expectation values to evaluate correction terms against the Gaussian weight. Our method allows us to examine the dependence of the path integral on the ordering parameter in addition to that on the splitting parameter. Although the effective potential for the fixed-energy amplitude derived by the method of Ref.\onlinecite{KleinertBook} is independent of the splitting parameter, our result is dependent on this parameter even in the path integral for the fixed-energy amplitude. The effective potential given by \eqref{eq:kleinert22} is obtained only for the symmetric setting for the regulating functions since we have an additional contribution proportional to $(1-2\lambda)^{2}$. As we have shown in section 4, the proof of the equivalence of the radial Coulomb and the radial oscillator path integrals in our formulation is based on this setting. Hence the $\lambda$-dependent correction term to the effective potential will cause changes in angular momenta for both of these systems. As a consequence the order of the modified Bessel function becomes $\lambda$-dependent to break the connection to the angular decomposition. To avoid this as well as to make $H_{E}$ be Hermitian, we should set $\lambda=1/2$.  The symmetric setting will therefore be preferred from these view points.

Since our method allows us to formulate path integrals in an arbitrary ordering prescription, its advantage in application will be evident. If we view the problem we have solved in this paper from a different angle, we may regard the DK Hamiltonian $H_{E}$ as the one for a system which possesses a position dependent mass. The technique that converts the non-trivial kinetic term into the effective potential may be therefore useful for solving path integrals for such systems.
Applications to path integrals in curved spaces in higher dimensions may also be possible by making a suitable extension if needed.

In the application to radial path integrals, we have introduced two sorts of completeness; one is formed by the radial plane wave and the other has been introduced to be an eigenfunction of the free Hamiltonian \eqref{eq:bessel01}. Despite the different constructions, the resulting path integrals coincide to each other to convince us the validity of these two formulations. Furthermore, the successful result in this application exhibits the usefulness of our method, which assumes the order estimation by regarding $\Delta x$, $\Delta q$, etc. as $O(\sqrt{\epsilon\,})$, even for systems defined on a half line. The same will be true even for a system whose domain is restricted to a finite region provided that the system does not possess the periodicity. To aim applications for such systems we will need a generalization of our method to implement the periodicity in a suitable way.

\appendix
\section{Conversion of a non-trivial kinetic term into the effective potential}
We show here the essence of our method of evaluating higher order terms in a series of $\Delta q$ in the exponent of a time sliced path integral.
For our aim, the relevant part of the time sliced path integral \eqref{eq:kernel-a2} for the Feynman kernel is given by
\begin{equation}
\label{eq:appa01}
K=\left(\frac{\ 1\ }{\ 2\pi\epsilon\ }\right)^{N/2}
\int_{-\infty}^{\infty}\!\prod_{i=1}^{N-1}dq_{i}\,
\exp\left[-\sum_{j=1}^{N}\frac{\ 1\ }{\ 2\epsilon\ }
\{\Delta q_{j}R(q^{(\alpha)}_{j},\Delta q_{j})\}^{2}\right],\quad
\epsilon=\frac{\ \hbar\epsilon_{\sigma}\ }{\ m\ }.
\end{equation}
Here $R(q^{(\alpha)}_{j},\Delta q_{j})$ is given by
\begin{equation}
\label{eq:appa02}
R(q^{(\alpha)}_{j},\Delta q_{j})=1+a_{2}(q^{(\alpha)}_{j})\Delta q_{j}+
a_{3}(q^{(\alpha)}_{j})(\Delta q_{j})^{2}+\cdots
\end{equation}
in the power series of $\Delta q_{j}$.
If we define
\begin{equation}
\label{eq:appa04}
q{'}_{j}\equiv q_{j}+\sum_{k=1}^{j}
\{R(q^{(\alpha)}_{k},\Delta q_{k})-1\}
\Delta q_{k}\ 
(j=1,\,2,\,\dots,\,N),\quad
q{'}_{0}\equiv q_{0},
\end{equation}
there holds
\begin{equation}
\label{eq:appa05}
\Delta q{'}_{j}\equiv q{'}_{j}-q{'}_{j-1}=
\Delta q_{j}R(q^{(\alpha)}_{j},\Delta q_{j}),
\end{equation}
by which we can rewrite the exponent of \eqref{eq:appa01} as a sum of $(\Delta q{'}_{j})^{2}$.

By solving the relation between $\Delta q_{j}$ and $\Delta q{'}_{j}$ with respect to $\Delta q_{j}$, we may find a series expansion
\begin{equation}
\label{eq:appa06}
\Delta q_{j}\equiv\Delta q{'}_{j}+
\tilde{a}_{2}(q{'}^{(\alpha)}_{j})
(\Delta q{'}_{j})^{2}+
\tilde{a}_{3}(q{'}^{(\alpha)}_{j})(\Delta q{'}_{j})^{3}+\cdots
\end{equation}
as the inverse of \eqref{eq:appa05}. Since $q^{(\alpha)}_{j}=q{'}^{(\alpha)}_{j}+O((\Delta q{'}_{j})^{2})$,
substitution of the above series into Eq.\eqref{eq:appa05} yields
\begin{equation}
\begin{aligned}
\Delta q{'}_{j}=&
\Delta q{'}_{j}+
\left\{\tilde{a}_{2}(q{'}^{(\alpha)}_{j})+a_{2}(q{'}^{(\alpha)}_{j})\right\}
(\Delta q{'}_{j})^{2}+\\
&
\left\{\tilde{a}_{3}(q{'}^{(\alpha)}_{j})+a_{3}(q{'}^{(\alpha)}_{j})+
2\tilde{a}_{2}(q{'}^{(\alpha)}_{j})a_{2}(q{'}^{(\alpha)}_{j})\right\}
(\Delta q{'}_{j})^{3}+\cdots.
\end{aligned}
\end{equation}
We thus obtain
\begin{equation}
\tilde{a}_{2}(q{'}^{(\alpha)}_{j})=-a_{2}(q{'}^{(\alpha)}_{j}),\quad
\tilde{a}_{3}(q{'}^{(\alpha)}_{j})=-a_{3}(q{'}^{(\alpha)}_{j})+
2\{a_{2}(q{'}^{(\alpha)}_{j})\}^{2},\quad\dots
\end{equation}
to find that $\Delta q_{j}$ is expressed as
\begin{equation}
\Delta q_{j}=\Delta q{'}_{j}-
a_{2}(q{'}^{(\alpha)}_{j})
(\Delta q{'}_{j})^{2}-
\{a_{3}(q{'}^{(\alpha)}_{j})-
2\{a_{2}(q{'}^{(\alpha)}_{j})\}^{2}\}(\Delta q{'}_{j})^{3}+\cdots
\end{equation}
in terms of $\Delta q{'}_{j}$. Therefore $q_{j}$ can be expressed in terms of $q{'}_{k}$'s as
\begin{equation}
q_{j}=q{'}_{j}-\sum_{k=1}^{j}\left[
a_{2}(q{'}^{(\alpha)}_{k})
(\Delta q{'}_{k})^{2}-
\left\{a_{3}(q{'}^{(\alpha)}_{k})-
2\{a_{2}(q{'}^{(\alpha)}_{k})\}^{2}\right\}(\Delta q{'}_{k})^{3}+\cdots\right].
\end{equation}

The derivative of $q_{j}$ with respect to $q{'}_{j}$ is given by
\begin{equation}
\frac{\ \partial q_{j}\ }{\ \partial q{'}_{j}\ }=1-2a_{2}(q{'}^{(\alpha)}_{j})
\Delta q{'}_{j}-3\left\{a_{3}(q{'}^{(\alpha)}_{j})-
2\{a_{2}(q{'}^{(\alpha)}_{j})\}^{2}+\frac{\ 1\ }{\ 3\ }
(1-\alpha)a_{2}{'}(q{'}^{(\alpha)}_{j})\right\}(\Delta q{'}_{j})^{2}+\cdots.
\end{equation}
For $k<j$, we get
\begin{equation}
\begin{aligned}
&\frac{\ \partial q_{j}\ }{\ \partial q{'}_{k}\ }=-2a_{2}(q{'}^{(\alpha)}_{k})
\Delta q{'}_{k}-3\left\{a_{3}(q{'}^{(\alpha)}_{k})-
2\{a_{2}(q{'}^{(\alpha)}_{k})\}^{2}+\frac{\ 1\ }{\ 3\ }
(1-\alpha)a_{2}{'}(q{'}^{(\alpha)}_{k})\right\}(\Delta q{'}_{k})^{2}\\
&+2a_{2}(q{'}^{(\alpha)}_{k+1})
\Delta q{'}_{k+1}+3\left\{a_{3}(q{'}^{(\alpha)}_{k+1})-
2\{a_{2}(q{'}^{(\alpha)}_{k+1})\}^{2}+\frac{\ 1\ }{\ 3\ }
\alpha a_{2}{'}(q{'}^{(\alpha)}_{k+1})\right\}(\Delta q{'}_{k+1})^{2}+\cdots
\end{aligned}
\end{equation}
and for $k>j$
\begin{equation}
\frac{\ \partial q_{j}\ }{\ \partial q{'}_{k}\ }=0.
\end{equation}
We thus see the matrix $J$ defined by
\begin{equation}
J_{ij}\equiv\frac{\ \partial q_{i}\ }{\ \partial q{'}_{j}\ }
\end{equation}
is triangular.
Therefore its determinant is given by the product of its diagonal elements:
\begin{equation}
\label{eq:jacobian01}
\begin{aligned}
\det J
=&\prod_{j=1}^{N-1}\left[\vphantom{\frac{1}{3}}
1-2a_{2}(q{'}^{(\alpha)}_{j})\Delta q{'}_{j}-\right.\\
&\left.
3\left\{a_{3}(q{'}^{(\alpha)}_{j})-
2\{a_{2}(q{'}^{(\alpha)}_{j})\}^{2}+\frac{\ 1\ }{\ 3\ }
(1-\alpha)a_{2}{'}(q{'}^{(\alpha)}_{j})\right\}(\Delta q{'}_{j})^{2}+\cdots
\right].
\end{aligned}
\end{equation}
This is the Jacobian for the change of variables from $q_{1},\,q_{2},\,\dots,\,q_{N-1}$ to
$q{'}_{1},\,q{'}_{2},\,\dots,\,q{'}_{N-1}$.
By taking these into account, we obtain
\begin{equation}
K=\left(\frac{\ 1\ }{\ 2\pi\epsilon\ }\right)^{N/2}
\int_{-\infty}^{\infty}\!\prod_{i=1}^{N-1}dq{'}_{i}\,\det J
\exp\left[-\sum_{j=1}^{N}\frac{\ 1\ }{\ 2\epsilon\ }
(\Delta q{'}_{j})^{2}\right].
\end{equation}

Due to the  Gaussian factor, we may treat $\Delta q{'}_{j}$ to be $O(\sqrt{\epsilon\,})$ for sufficiently small $\epsilon$.
Then by arranging $\partial q_{j}/\partial q{'}_{j}$ as
\begin{equation}
\begin{aligned}
&1-2a_{2}(q{'}^{(\alpha)}_{j})\Delta q{'}_{j}-
3\left\{a_{3}(q{'}^{(\alpha)}_{j})-
2\{a_{2}(q{'}^{(\alpha)}_{j})\}^{2}+\frac{\ 1\ }{\ 3\ }
(1-\alpha)a_{2}{'}(q{'}^{(\alpha)}_{j})\right\}(\Delta q{'}_{j})^{2}+\cdots\\
=&
\left[1-2a_{2}(q{'}^{(\alpha)}_{j})\Delta q{'}_{j}
+2\{a_{2}(q{'}^{(\alpha)}_{j})\Delta q{'}_{j}\}^{2}+\cdots\right]\times\\
&\left[1-
\left\{3a_{3}(q{'}^{(\alpha)}_{j})-
4\{a_{2}(q{'}^{(\alpha)}_{j})\}^{2}+
(1-\alpha)a_{2}{'}(q{'}^{(\alpha)}_{j})\right\}(\Delta q{'}_{j})^{2}+\cdots
\right],
\end{aligned}
\end{equation}
we see
\begin{equation}
\begin{aligned}
&1-2a_{2}(q{'}^{(\alpha)}_{j})\Delta q{'}_{j}-
3\left\{a_{3}(q{'}^{(\alpha)}_{j})-
2\{a_{2}(q{'}^{(\alpha)}_{j})\}^{2}+\frac{\ 1\ }{\ 3\ }
(1-\alpha)a_{2}{'}(q{'}^{(\alpha)}_{j})\right\}(\Delta q{'}_{j})^{2}+\cdots\\
=&
\exp\left[
-2a_{2}(q{'}^{(\alpha)}_{j})\Delta q{'}_{j}-
\left\{3a_{3}(q{'}^{(\alpha)}_{j})-
4\{a_{2}(q{'}^{(\alpha)}_{j})\}^{2}+
(1-\alpha)a_{2}{'}(q{'}^{(\alpha)}_{j})\right\}(\Delta q{'}_{j})^{2}
\right]
\end{aligned}
\end{equation}
to be correct up to $O(\epsilon)$ in the exponent.
We thus obtain
\begin{equation}
\begin{aligned}
&K=\left(\frac{\ 1\ }{\ 2\pi\epsilon\ }\right)^{N/2}
\int_{-\infty}^{\infty}\!\prod_{i=1}^{N-1}dq{'}_{i}\,
\exp\left[-\sum_{j=1}^{N}\frac{\ 1\ }{\ 2\epsilon\ }
(\Delta q{'}_{j})^{2}\right]\\
&\times
\exp\left[
-\sum_{j=1}^{N}\left\{
2a_{2}(q{'}^{(\alpha)}_{j})\Delta q{'}_{j}+
\left\{3a_{3}(q{'}^{(\alpha)}_{j})-
4\{a_{2}(q{'}^{(\alpha)}_{j})\}^{2}+
(1-\alpha)a_{2}{'}(q{'}^{(\alpha)}_{j})\right\}
(\Delta q{'}_{j})^{2}\right\}\right]
\end{aligned}
\end{equation}
by discarding irrelevant terms in the limit $\epsilon\to0$.
The $N$-th term of the sum in the second line above was absent in the original Jacobian \eqref{eq:jacobian01},
but it will disappear in the limit $\epsilon\to0$. Therefore we are allowed to add this to the sum without changing the result.

In order to convert the terms generated by the Jacobian into the effective potential, we complete the square in the exponent as
\begin{equation}
\begin{aligned}
&\frac{\ 1\ }{\ 2\epsilon\ }(\Delta q{'}_{j})^{2}+
2a_{2}(q{'}^{(\alpha)}_{j})\Delta q{'}_{j}+
\left\{3a_{3}(q{'}^{(\alpha)}_{j})-
4\{a_{2}(q{'}^{(\alpha)}_{j})\}^{2}+
(1-\alpha)a_{2}{'}(q{'}^{(\alpha)}_{j})\right\}
(\Delta q{'}_{j})^{2}\\
=&
\frac{\ 1\ }{\ 2\epsilon\ }\left\{
1+\epsilon C_{2}(q{'}^{(\alpha)}_{j})\right\}
\left\{\Delta q{'}_{j}+
\frac{\ \epsilon C_{1}(q{'}^{(\alpha)}_{j})\ }
{\ 1+\epsilon C_{2}(q{'}^{(\alpha)}_{j})\ }\right\}^{2}-
\frac{\ \epsilon\ }{\ 2\ }\{C_{1}(q{'}^{(\alpha)}_{j})\}^{2}+O(\epsilon^{2}),
\end{aligned}
\end{equation}
in which we have defined
\begin{equation}
C_{1}(q)\equiv
2a_{2}(q),\quad
C_{2}(q)\equiv
2\left\{3a_{3}(q)-
4\{a_{2}(q)\}^{2}+
(1-\alpha)a_{2}{'}(q)\right\}.
\end{equation}
If we can find suitable changes of variables to yield
\begin{equation}
\label{eq:square01}
\frac{\ 1\ }{\ 2\epsilon\ }\left\{
1+\epsilon C_{2}(q{'}^{(\alpha)}_{j})\right\}
\left\{\Delta q{'}_{j}+
\frac{\ \epsilon C_{1}(q{'}^{(\alpha)}_{j})\ }
{\ 1+\epsilon C_{2}(q{'}^{(\alpha)}_{j})\ }\right\}^{2}=
\frac{\ 1\ }{\ 2\epsilon\ }(\Delta\xi_{j})^{2},
\end{equation}
we will obtain a time sliced path integral accompanied by an effective potential in addition to the standard kinetic term. To this aim, we define
\begin{equation}
\eta_{j}\equiv q{'}_{j}+\sum_{k=1}^{j}
\frac{\ \epsilon C_{1}(q{'}^{(\alpha)}_{k})\ }
{\ 1+\epsilon C_{2}(q{'}^{(\alpha)}_{k})\ }
\end{equation}
to find
\begin{equation}
\Delta\eta_{j}=\Delta q{'}_{j}+
\frac{\ \epsilon C_{1}(q{'}^{(\alpha)}_{j})\ }
{\ 1+\epsilon C_{2}(q{'}^{(\alpha)}_{j})\ }.
\end{equation}
We then introduce $\xi_{j}$ with a new function $G(q)$ by
\begin{equation}
\xi_{j}\equiv\sqrt{1+\epsilon G(\eta_{j})\,}\eta_{j}.
\end{equation}
If the function $G(q)$ satisfies
\begin{equation}
\label{eq:DEQofG}
G(q)+G{'}(q)q=C_{2}(q),
\end{equation}
there holds
\begin{equation}
\Delta\xi_{j}=\sqrt{1+\epsilon C_{2}(\eta^{(\alpha)}_{j})\,}\Delta\eta_{j}
\end{equation}
and hence \eqref{eq:square01} by discarding irrelevant terms.
The Jacobian induced by these two steps of changing variables reads
\begin{equation}
\exp\left[
-\frac{\ \epsilon\ }{\ 2\ }\sum_{j=1}^{N}
\left\{C_{2}(\xi^{(\alpha)}_{j})+
2(1-\alpha)C_{1}{'}(\xi^{(\alpha)}_{j})
\right\}
\right].
\end{equation}
In this way, we obtain a time sliced path integral
\begin{equation}
\label{eq:kernel01}
\begin{aligned}
&K=\left(\frac{\ 1\ }{\ 2\pi\epsilon\ }\right)^{N/2}
\int_{-\infty}^{\infty}\!\prod_{i=1}^{N-1}d\xi_{i}\,
\exp\left[-\sum_{j=1}^{N}\frac{\ 1\ }{\ 2\epsilon\ }
(\Delta \xi_{j})^{2}\right]\\
&\times
\exp\left[
-\frac{\ \epsilon\ }{\ 2\ }\sum_{j=1}^{N}\left\{
\{C_{1}(\xi^{(\alpha)}_{j})\}^{2}+C_{2}(\xi^{(\alpha)}_{j})+
2(1-\alpha)C_{1}{'}(\xi^{(\alpha)}_{j})
\right\}\right]
\end{aligned}
\end{equation}
which is equivalent to Eq.\eqref{eq:appa01}.
In obtaining the above, we have taken the fact  $q{'}_{j}^{(\alpha)}=\xi_{j}^{(\alpha)}+O(\epsilon)$ into account
to arrange the exponent by the substitution $q{'}_{j}^{(\alpha)}\to\xi_{j}^{(\alpha)}$. Also $C_{2}(\xi_{j})$ in the sum has been replaced by $C_{2}(\xi^{(\alpha)}_{j})$ because of the existence of $\epsilon$ in front of the sum.
We thus find that higher order terms with respect to $\Delta q_{j}$ in the original path integral \eqref{eq:appa01} can be converted into the effective potential, given by
\begin{equation}
\label{eq:Veff01}
\begin{aligned}
V_{\mathrm{eff}}(q)=&\frac{\ 1\ }{\ 2\ }
\left\{\{C_{1}(q)\}^{2}+C_{2}(q)+
2(1-\alpha)C_{1}{'}(q)
\right\}\\
=&
3\{a_{3}(q)-\alpha a_{2}{'}(q)\}-2\{a_{2}(q)\}^{2}+3a_{2}{'}(q),
\end{aligned}
\end{equation}
in \eqref{eq:kernel01} which has the kinetic term in a standard form.

We have almost sufficiently achieved our aim of this appendix excepting to check the dependence of the path integral \eqref{eq:kernel01} on the ordering parameter $\alpha$ which has been introduced through $R(q^{(\alpha)}_{j},\Delta q_{j})$ in \eqref{eq:appa01}. If we shift the ordering parameter from $\alpha$ to $\alpha{'}=\alpha+\delta\alpha$, $q^{(\alpha)}_{j}$ is shifted to $q^{(\alpha{'})}_{j}=q^{(\alpha)}_{j}-\delta\alpha\Delta q_{j}$ and a corresponding change in $R(q^{(\alpha)}_{j},\Delta q_{j})$ reads
\begin{equation}
\label{eq:alpha01}
\begin{aligned}
R(q^{(\alpha{'})}_{j},\Delta q_{j})=&1+a_{2}(q^{(\alpha{'})}_{j})\Delta q_{j}+
a_{3}(q^{(\alpha{'})}_{j})(\Delta q_{j})^{2}+\cdots\\
=&1+a_{2}(q^{(\alpha)}_{j})\Delta q_{j}+
\left\{a_{3}(q^{(\alpha)}_{j})-\delta\alpha a_{2}{'}(q^{(\alpha)}_{j})\right\}
(\Delta q_{j})^{2}+\cdots.
\end{aligned}
\end{equation}
We thus find that only $a_{3}(q^{(\alpha)}_{j})$ alone in the relevant terms will be transformed into $a_{3}(q^{(\alpha)}_{j})-\delta\alpha a_{2}{'}(q^{(\alpha)}_{j})$ by this change in the ordering parameter.
The structure in the first term of the second line of \eqref{eq:Veff01}
clearly exhibits this property: possible change in the time sliced path integral \eqref{eq:kernel01} is restricted to the shift $\alpha\to\alpha+\delta\alpha$ in the coefficients of the effective potential because the change $\xi^{(\alpha)}_{j}$ to $\xi^{(\alpha{'})}_{j}$ generates just irrelevant terms due to the existence of $\epsilon$. If, by some reason, there exists a nice relation between $a_{2}{'}(q)$ and $a_{3}(q)$, the cancellation in the $\alpha$-dependent terms in the effective potential may happen to occur. This is the case for the system treated in the text.

\end{document}